\documentclass[fleqn,usenatbib]{mnras}

\usepackage{newtxtext,newtxmath}
\usepackage[T1]{fontenc}

\usepackage{amsmath}
\usepackage{graphicx}
 
 \DeclareRobustCommand{\VAN}[3]{#2}
\let\VANthebibliography\thebibliography
\def\thebibliography{\DeclareRobustCommand{\VAN}[3]{##3}\VANthebibliography}
 
\newcommand{\lowz}{LOWZ}
\newcommand{\cmass}{CMASS}

\newcommand{\cii}{[C{\sc\,II}]}
\newcommand{\ci}{[C{\sc\,I}]}

\newcommand*\mean[1]{\bar{#1}}
\newcommand{\pcl}{pseudo-$C_{\ell}$}
\newcommand{\FF}{FIRAS${\times}$FIRAS}
\newcommand{\BB}{BOSS${\times}$BOSS}

\newcommand{\FB}{FIRAS${\times}$BOSS}
\newcommand{\PB}{Planck${\times}$QSO}
\newcommand{\PIB}{PIXIE${\times}$CMASS}
\newcommand{\FLZ}{FIRAS${\times}$LOWZ}
\newcommand{\FC}{FIRAS${\times}$CMASS}

\title[FIRASxBOSS \cii\ IM Constraints]{Constraining low redshift \cii\ Emission by Cross-Correlating FIRAS and BOSS Data}

\author[C. J. Anderson et al.]{
C. J. Anderson$^{1}$\thanks{E-mail: christopher.j.anderson@nasa.gov}, E. R. Switzer$^{1}$,
P.~C.~Breysse$^{2}$\\
$^{1}$NASA Goddard Space Flight Center, Greenbelt, MD 20771, USA\\
$^{2}$Center for Cosmology and Particle Physics, Department of Physics, New York University, 726 Broadway, New York, NY, 10003, U.S.A.}

\date{Accepted XXX. Received YYY; in original form ZZZ}

\pubyear{2022}

\begin{document}
\label{firstpage}
\pagerange{\pageref{firstpage}--\pageref{lastpage}}
\maketitle

\begin{abstract}
We perform a tomographic cross-correlation analysis of archival FIRAS data and the BOSS galaxy redshift survey
to constrain the amplitude of \cii\ $^2P_{3/2}\rightarrow$ $^2P_{1/2}$ fine structure emission. Our analysis employs spherical harmonic tomography (SHT), which is based on the angular cross-power spectrum between FIRAS maps and BOSS galaxy over-densities at each pair of redshift bins, over a redshift range of $0.24<z<0.69$. We develop the SHT approach for intensity mapping, where it has several advantages over existing power spectral estimators. Our analysis constrains the product of the \cii\ bias and \cii\ specific intensity, $b_{\rm \cii}I_{\rm \cii}$, to be $<0.31$\,MJy/sr at $z {\approx} 0.35$ and $<0.28$\,MJy/sr at $z {\approx} 0.57$ at $95\%$ confidence. These limits are consistent with most current models of the \cii\ signal, as well as with higher-redshift \cii\ cross-power spectrum measurements from the Planck satellite and BOSS quasars.  We also show that our analysis, if applied to data from a more sensitive instrument such as the proposed PIXIE satellite, can detect pessimistic \cii\ models at high significance.
\end{abstract} 

\begin{keywords}
galaxies: evolution -- methods: data analysis -- infrared: general
\end{keywords}

\section{Introduction}
\label{sec:introduction}

Atomic and molecular line emissions are powerful diagnostic tracers of the evolution of star formation over cosmic time. For instance, line emission studies may help reveal the causes of the significant decline in star formation rate after its peak at $z {\sim} 2$, inferred from observations of optical and infrared continuum radiation \citep{Madau2014}. One of the most useful lines for understanding the context of star formation is the \cii\ $^2P_{3/2}\rightarrow$ $^2P_{1/2}$ fine structure line at $1901$\,GHz, which is the brightest cooling line in the far-infrared (FIR) spectrum, typically accounting for 0.1$\%$ to 1$\%$ of FIR energy \citep{1991ApJ...373..423S, malhotra1997infrared, diaz-santos:2017}. The bulk of this \cii\ emission is expected to come from photodissociation regions (PDRs) on the edges of molecular gas clouds \citep{2010ApJ...724..957S}. This association suggests that \cii\ emission can trace the molecular gas available to fuel star formation. A study of $z\sim2$ galaxies detected by ALMA bore this out, finding that \cii\ emission is linearly related to molecular gas content \citep{2018MNRAS.481.1976Z}. Measurements of \cii\ emission and other lines (such as \ci\ and the rotational transitions of CO) can be used in concert with galaxy formation models to constrain the physical properties of star-forming regions within galaxies \citep{2019MNRAS.482.4906P, 2021ApJ...911..132Y}. 

Studies of line emission from individual galaxies (e.g., \citet{2013ARA&A..51..105C, 2017ApJ...834...36H, 2018MNRAS.481.1976Z}) provide important insights into the \cii\ luminosity function. However, they are subject to sample variance in small survey regions and are limited to galaxies above a brightness threshold, so they may not capture the cosmic average of the conditions of star formation. A complementary technique to individual galaxy studies is the technique of intensity mapping, which aims to map large-scale structure by detecting the aggregate redshifted line emission without cataloging individual sources \citep{1979MNRAS.188..791H, 1990MNRAS.247..510S, 1997ApJ...475..429M, 1999ApJ...512..547S, 2008MNRAS.383.1195W, 2008PhRvL.100i1303C, 2011JCAP...08..010V, kovetz2017line, 2019BAAS...51c.101K}. Some of the advantages of intensity mapping are that it captures all sources of emission rather than a biased sample of only the brightest sources. It also puts significantly lower requirements on telescope size, as the angular resolution need not be sufficient for individual source detection. Similarly, because individual objects do not need to be detected at a high signal-to-noise ratio, intensity mapping surveys can quickly cover large cosmic volumes, providing a complete census of emitting gas.

Intensity mapping has developed rapidly in the past fifteen years. Pathfinding intensity mapping surveys have used pre-existing telescopes to detect aggregate emission from the 21-cm line of neutral hydrogen (HI) \citep{2009MNRAS.394L...6P, 2010Natur.466..463C, 2013ApJ...763L..20M, 2013MNRAS.434L..46S, 2017MNRAS.464.4938W, 2018MNRAS.476.3382A, 2021arXiv210204946W} via cross-correlation with optical galaxy surveys. Several dedicated 21-cm intensity mapping experiments are now underway, targeting both the epoch of reionization (e.g., LOFAR, \cite{2013A&A...556A...2V}, and SKA precursors HERA and MWA, \cite{DeBoer_2017,2018MNRAS.481.5034M}) as well as the era of dark energy dominance (e.g., CHIME, \cite{2014SPIE.9145E..22B}, Tianlai, \cite{2012IJMPS..12..256C}, HIRAX, \cite{2021arXiv210913755C}, SKA precursor MeerKAT, \cite{2021MNRAS.505.3698W}, and BINGO, \cite{2021arXiv210701633A}). The initial focus on the 21-cm line has expanded to include \cii, the rotational lines of CO, Lyman $\alpha$, H$\alpha$, H$\beta$, and more \citep{2019BAAS...51c.101K}. 

The COPPS survey has utilized the Sunyaev-Zel'dovich Array (SZA) to make tentative detections of CO emission at $z\sim2.6$ in cross-correlation with spectroscopic galaxy surveys \citep{2021arXiv211002239K} and auto-correlation in the shot-noise regime \citep{Keating_2016}. A similar CO shot-noise detection was made with Millimeter-wave Intensity Mapping Experiment (mmIME), using ALMA and ACA facilities \citep{Keating_2020}. Among a new generation of dedicated ground-based intensity mapping instruments targeting CO and \cii\ are COMAP \citep{2021arXiv211105927C}, designed to measure CO(1-0) from $2.4<z<3.4$ and CO(2-1) at $z=6-8$ , TIME \citep{2014SPIE.9153E..1WC}, targeting \cii\ emission from the epoch of reionization and CO from $0<z<2$, CONCERTO \citep{2020arXiv200714246T}, focusing on \cii\ emission from the epoch of reionization, the CCAT-prime receiver for the FYST \citep{2021arXiv210710364C}, which targets \cii\ at $3.5<z<8$ and [OIII] at $z>7$, and SPT-SLIM \citep{2022JLTP..tmp...61K}, which aims to use the South Pole Telescope to target several CO transitions from $0.3<z<2.8$. Two NASA balloon experiments designed to measure line emission in the far-infrared (FIR) are TIM \citep{2020arXiv200914340V}, focusing on \cii\ at $0.52<z<1.67$, and EXCLAIM \citep{2021arXiv210111734C}, focusing on CO and \cii\ in windows from $0<z<3.5$. The NASA MIDEX-class satellite mission, SPHEREx, will produce intensity maps of multiple lines, including H$\alpha$, H$\beta$, [OII], and [OIII] at $z<5$ \citep{2017ApJ...835..273G}.

The brightness of the \cii\ line makes it an excellent candidate for the technique of line intensity mapping. \cite{pullen2018search} recently demonstrated the promise of \cii\ intensity mapping through an analysis of the cross-correlation between the angular distribution of high-redshift ($z {\sim} 2.6$) quasars in the BOSS survey and the $353$, $545$, and $857$\,GHz Planck maps. They found a cross-correlation exceeding the expected thermal continuum at $545$\,GHz, consistent with \cii\ emission correlated with BOSS quasars. A refinement of the analysis \citep{yang2019evidence} increased the result's significance. Still, the authors caution that greater spectral resolution is required to verify that the excess cross-correlation is explained by \cii\ emission rather than the redshift evolution of the correlated continuum emission.

\citet{2017ApJ...838...82S} describes how future instruments for measuring spectral distortions in the cosmic microwave background (CMB) can be employed for intensity mapping. To anticipate the capabilities of future measurements, we use data from the COBE-FIRAS instrument \citep{fixsen1994calibration}, in cross-correlation with the BOSS CMASS and LOWZ galaxy catalogs, to make the first tomographic intensity mapping constraint on \cii\ emission. 
The FIRAS instrument \citep{fixsen1994calibration} was designed to precisely measure the spectrum of the CMB, dust, and line emission from the Milky Way. It covers a broad frequency range from $30$\,GHz to $2910$\,GHz\ with $13.6$\,GHz spectral channels. This, and the fact that FIRAS' frequency range conveniently overlaps the well-sampled LOWZ and CMASS galaxy catalogs, make \FB\, a natural candidate for a tomographic \cii\, cross-correlation analysis. 

Compared to the Planck data set used by \citet{pullen2018search}, the FIRAS data set has much higher thermal noise per pixel and much lower angular resolution (the FIRAS beam has nearly a 7-degree full-width-half-maximum). 
At first glance, it seems that these limitations give \FB\ no hope of approaching the sensitivity of \PB, but we achieve error bars that are only about two times larger. One reason is that the effective number of independent modes scales with the number of redshift bins. We use 14 redshift bins for \FC\ and 16 for \FLZ, whereas \PB\ only has one channel with correlated \cii\ signal. However, the number of modes also depends on the range of measurable angular scales, and \PB\ more than makes up for its lack of redshift resolution with its much larger range of observable angular modes. Counteracting this, \FB\ sees a larger cosmological signal due to structure growth at lower redshifts, and, critically, the CMASS and LOWZ galaxy catalogs are well-sampled enough that shot noise is subdominant at the redshifts and angular scales we consider. The BOSS quasar sample used in \PB, on the other hand, is dominated by shot noise due to the sparser sampling.

\subsection{Spherical harmonic tomography}

We cross-correlate FIRAS data with cosmological overdensity inferred from the BOSS spectroscopic galaxy redshift survey to search for extragalactic \cii\ emission. Two competing effects make the BOSS \cmass\ ($0.41<z<0.75$) and \lowz\ ($0.06<z<0.49$) especially well-tuned for this goal. The growth of large-scale structure and geometric factors yield increasing density contrast on large angular scales where FIRAS is most sensitive. However, by $z<0.25$ ($\nu > 1500$\,GHz), the FIRAS noise rises dramatically. For our analysis, we bin the \cmass\ and \lowz\ galaxy maps into redshift slices that correspond to the FIRAS frequency channels for the \cii\ line. We then use the technique of spherical harmonic tomography (SHT) (see, for example, \cite{asorey2012recovering, nicola2014three, 2015A&A...578A..10L}) to compute the angular cross-power spectrum, $C_{\ell}^{\times}(z,z')$, between the FIRAS maps and BOSS galaxy over-densities at each pair of redshifts. 

With a fine enough binning in redshift, SHT captures the complete information available in the power spectrum \citep{asorey2012recovering}, and it is well-matched to large area surveys such as BOSS, where a flat-sky approximation would introduce significant distortions. In order to fit our data to a model, the expected angular clustering signal, $C_{\ell}^{\delta}(z,z')$, must be computed by integrating a 3D power spectrum model over highly oscillatory Bessel functions (details included in Section\,\ref{subsec:dark_matter_model}). Standard anisotropy codes, such as CAMB \citep{2011ascl.soft02026L} and CLASS \citep{2013JCAP...11..044D, 2014JCAP...01..042D}, include methods for performing this integration (with or without the Limber approximation). 

We describe the analysis over several sections. Section\,\ref{sec:C_ell_Analysis} motivates the $C_{\ell}(z,z')$ statistic, explains how to use the \pcl\ technique to compute $C_{\ell}(z,z')$ for surveys with partial sky coverage, and describes the likelihood function we use for parameter estimation with $C_{\ell}(z,z')$.
Section\,\ref{sec:FIRAS_and_BOSS_data} describes the FIRAS dataset, viewed as a \cii\ intensity map, and the BOSS \cmass\ and \lowz\ data sets, viewed as binned galaxy over-density maps.
Section \ref{sec:signal model} describes our dark matter model and its use in fitting the \cii\ and CIB amplitude to our \FB\ data. It also describes how we model the \FB\ covariance by fitting parametric models to the \BB\ and \FF\ data.
Section\,\ref{sec:Discussion} discusses our \cii\ constraints and how they relate to other measurements and astrophysical models. We conclude in Section\,\ref{sec:Conclusion}. 

\section{Parameter Estimation with the \texorpdfstring{$C_{\ell}(z,z')$}{Cl(z,z')} Statistic}
\label{sec:C_ell_Analysis}

\subsection{Motivating the estimator}
\label{subsec:motivating_estimator}

We perform our analysis with Spherical Harmonic Tomography (SHT), a two-point statistic wherein each redshift slice of data is decomposed into spherical harmonics, and the angular power spectrum, $C_{\ell}(z,z')$, is calculated between each pair of redshift slices. SHT captures the complete information available in the more typically used three-dimensional power spectrum statistic, $P(k)$ \citep{asorey2012recovering}. The SHT technique has already been used in several clustering analyses and forecasts for galaxy surveys (e.g. \cite{PhysRevLett.106.241301, 2013MNRAS.435.1857L, 2018MNRAS.476.1050B, loureiro2019cosmological,  2019JCAP...08..037X,
2019JCAP...12..028F, 2020JCAP...03..044N, 
2020JCAP...09..054V,
2021arXiv210700026T}). Among these, a recent analysis of BOSS CMASS and LOWZ clustering using Spherical Harmonic Tomography \citep{loureiro2019cosmological} found equivalent or better constraints on cosmological parameters compared to standard power spectrum analysis techniques. As the set of two-point cross-correlations, the SHT contains all information in the data cubes that is statistically isotropic and Gaussian.

SHT has some inherent geometrical advantages, especially for large-angle and deep surveys. A significant advantage is that the spherical coordinates apply to wide-angle surveys like BOSS without any flat-sky approximation. A traditional $P(k)$ analysis relies on the flat-sky approximation to distinguish between transverse and line-of-sight modes, which is critical for accurately representing redshift space distortions (RSDs) and distinguishing continuum foregrounds from line signal. By contrast, in the SHT formalism, both foregrounds and linear redshift space distortions take exact, simple forms. This geometric advantage is shared with the related analysis technique of spherical Fourier-Bessel (SFB) decomposition \citep{1995MNRAS.272..885F, 1995MNRAS.275..483H, 2012A&A...540A.115R, 2012A&A...540A..60L, 2013PhRvD..88b3502Y,
2015A&A...578A..10L,
2016ApJ...833..242L, PhysRevD.104.123548}, which, in addtion to spherical harmonic decomposition, involves a further Fourier-Bessel transform and data-compression along the line-of-sight. However, SFB does not share another important feature of SHT, which is its ability to capture redshift-dependent change over cosmological time in deep surveys. 
For deep surveys, structure growth and changes in star formation rate break the assumption of translational invariance in the line-of-sight direction, rendering the $P(k)$ or SFB statistic insufficient. However, since $C_{\ell}(z,z')$ does not compress data along the line-of-sight direction, it describes redshift evolution. A study \citep{2022arXiv220311095M} of simulated 21-cm data from the Epoch of Reionization (EoR) found that an implementation of the SHT technique, MAPS \citep{2018MNRAS.474.1390M, 2019MNRAS.483L.109M, 2020MNRAS.494.4043M},  obtains $\sim$2 times more stringent error bars on model parameters than techniques that fail to capture redshift evolution due to data compression along the line-of-sight, such as $P(k)$ or SFB. A final geometric advantage is that $C_{\ell}(z,z')$ describes the data in observing coordinates of angle and frequency (or, equivalently, redshift) rather than re-gridding the data onto cosmological distances in an assumed cosmological model. An MCMC likelihood analysis that constrains cosmological parameters would therefore not need to recompute the data statistic at each step, which in principle would be needed for a $P(k)$ or SFB analysis.

There are several practical advantages to parameter estimation with SHT. Due to the scan strategy, intensity mapping data generally have inhomogeneous noise and partial sky coverage in the angular direction. Multiplication by noise weights in real space couples transverse modes in Fourier space, which can be easily accounted for in SHT analysis, thanks to pre-existing work \citep{hivon2002master, tristram2005xspect} on the \pcl\ technique from CMB analysis. Intensity maps will also often have significant variation in the noise level in the frequency (line-of-sight) direction because of variations in spectrometer noise or chromatic contamination such as terrestrial radio frequency interference. Inhomogeneous noise in real space produces coupling of line-of-sight Fourier modes in analyses of $P(k)$. In contrast, because $C_{\ell}(z,z')$ does not perform any Fourier or Bessel transform in the redshift direction, no coupling occurs, and the noise can be expressed as a simple function of the redshift slice. Additionally, chromatic beam effects can be modeled per redshift slice without introducing any flat-sky approximation. 

In addition, SHT provides an avenue for self-consistently handling foregrounds and other continuum signals. In contrast, other approaches that remove foregrounds in map space before the two-point analysis must contend with signal loss \citep{Switzer:2015ria, 2018ApJ...868...26C}. With SHT, the covariance of $C_{\ell}(z,z')$ can be constructed to include information about foregrounds. The inverse covariance weights the data in the likelihood analysis for the line brightness, suppressing modes contaminated by foregrounds self-consistently with the parameter estimation. Additionally, the $C_\ell(z,z')$ model can include signal terms such as the cross-correlation of a galaxy redshift sample with the correlated continuum emission of the galaxies in the IM volume \citep{Serra:2014pva, pullen2018search, Switzer:2018tel}. 

SHT also introduces several new challenges to the analysis. First, it places significant requirements on memory for the computation of the likelihood. This is because the size of the covariance scales as $N_b^2 N_z^4$, where $N_z$ is the number of redshift bins and $N_b$ is the number of angular bins. For FIRAS, this is fairly manageable because we analyze only three angular bins and about 15 redshift slices for each of \FC\ and \FLZ. The size of the covariance will be more challenging for future instruments with higher spectral and angular resolution.

The visualization of $C_{\ell}(z,z')$ and its errors presents an additional challenge. The extra dimensionality of $C_{\ell}(z,z')$, which allows it to measure redshift evolution, also means that angular and redshift information cannot be shown in a single plot. We develop several approaches for displaying the data and describing the goodness of fit to high-dimensional data with complex covariance. Lastly, the evaluation of the $C_{\ell}(z,z')$ model is computationally expensive. Since the current generation of IM observations focuses on detection and measurements of line brightness \citep{2019BAAS...51c.101K}, this is not an issue. In this regime, the $C_{\ell}(z,z')$ model can be constructed from linear combinations of cosmological clustering and shot noise templates that need only be calculated once. In contrast, studies of large-scale structure acquire information from the nonlinear dependence of $C_{\ell}(z,z')$ on cosmological parameters, so can require expensive recalculation of the anisotropy. Recent work has accelerated the integrals that convert $P(k,z)$ to $C_{\ell}(z,z')$ without using the Limber approximation \citep{2017A&A...602A..72C, Schoneberg:2018fis}, making cosmological parameter estimation with $C_{\ell}(z,z')$ more practical.

\subsection{Computing \texorpdfstring{$C_{\ell}(z,z')$}{Cl(z,z')} with Incomplete Sky Coverage}
\label{subsec:Partial_sky}

Extensive work in CMB data analysis has developed an approach \citep{hivon2002master} for dealing with incomplete sky coverage and an approximate formula for the covariance it induces between angular scales \citep{tristram2005xspect}. The code package NaMaster \citep{Alonso:2018jzx,NaMaster} includes this full functionality, along with contaminant removal for polarized and unpolarized maps, on both large curved-sky maps and small maps, using a flat-sky approximation.  

For observed maps A and B with known, possibly different, inverse noise weights and beams, the angular power spectrum of the inverse noise weighted maps is the \pcl\ spectrum \citep{hivon2002master}, which we label $D_{\ell}$, following the notation of \cite{tristram2005xspect}. The \pcl\ spectrum is related to the true full-sky angular power spectrum $C_{\ell}$ by
\begin{equation}\label{eq:pseudocl}
D_{\ell} = \sum_{\ell'} M^{AB}_{\ell\ell'}\omega_{\ell'}^A\omega_{\ell'}^BC_{\ell'},
\end{equation}
where $\omega^A_{\ell'}$ is the product of the beam and pixel window function of map A, $\omega^B_{\ell'}$ is the product of the beam and pixel window function of map B, and $M^{AB}_{\ell \ell'}$ is the mixing matrix, computed via
\begin{equation}\label{eq:mixing_matrix}
M^{AB}_{\ell\ell'} = \frac{2\ell' + 1}{4 \pi} \sum_{\ell''}(2\ell''+1)\mathcal{W}^{AB}_{\ell''}\begin{pmatrix} \ell&\ell'&\ell'' \\ 0&0&0\end{pmatrix}^2,
\end{equation}
where $\mathcal{W}^{AB}_{\ell''}$ is the angular power spectrum of the inverse noise spatial weights of maps $A$ and $B$, and $\begin{pmatrix} \ell&\ell'&\ell'' \\ 0&0&0\end{pmatrix}$ represents a Wigner 3-j symbol. These formulas work for both auto and cross-correlations. The mixing matrix is not invertible with partial sky coverage, and it is convenient to define the binned \pcl\ spectrum and binned mixing matrix. Here, a range of multipoles $\ell$ are combined into a bandpower $b$, for which the mixing matrix of bandpowers $b$ and $b'$ is invertible using suitable binning. Define $B_{b\ell}$ as a binning matrix that averages blocks of consecutive $\ell$s into bins indexed by $b$ and define $B^u_{\ell b}$ as an equivalent unbinning matrix that assigns the average value of a bin $b$ to each $\ell$ within that bin. Then, define the binned quantities
\begin{equation}
\begin{split}\label{eq:binning_equation}
D_b &\equiv \sum_{\ell} B_{b\ell} D_{\ell}\\
M^{AB}_{b b'} &\equiv \sum_{\ell, \ell'} B_{b\ell}M^{AB}_{\ell\ell'}\omega^A_{\ell'}\omega^B_{\ell'}B^u_{\ell' b'}.
\end{split}
\end{equation}
If the bin size is wide enough, $M^{AB}_{b b'}$ is invertible even for partial sky coverage. One can then form an estimate, $\hat{C}_b$, of the binned angular power spectrum from the observed binned pseudo-$C_{b}$ spectrum, $\hat{D}_b$, via
\begin{equation}\label{eq:binned_mixing}
\hat{C}_b = \sum_{b'} \left( M^{AB}_{b b'}\right)^{-1} \hat{D}_{b'}.
\end{equation}
This procedure should only recover the true angular power spectrum if $C_{\ell}$ is constant within each bin $b$. Since this is not generally the case, to compare $\hat{C}_{b}$ to a model $C_{\ell}$, as we do in Section\,\ref{sec:signal model}, we always compute or simulate the expected binned angular power spectrum, $\tilde{C}_b$, which is related to the model via
\begin{equation}\label{eq:model_mixing_binning}
\tilde{C}_b \equiv \sum_{b'} \left(M^{AB}_{b b'}\right)^{-1} \sum_{\ell} B_{b' \ell} M^{AB}_{\ell \ell'} \omega^A_{\ell'}\omega^B_{\ell'} C_{\ell'}.
\end{equation}
If one assumes that the true $C_{\ell}(z,z')$ distribution is Gaussian, then the $b = b'$ terms of the covariance can be roughly approximated as
\begin{equation}
\begin{split}\label{eq:general_gaussian_variance}
    &\langle \Delta C_b^{A\times B}(z_1,z_2) \Delta C_b^{A\times B}(z_3,z_4) \rangle \approx \\ &\frac{1}{f_{\rm sky} \Delta \ell (2\ell + 1)} [ C_b^{A\times A}(z_1,z_3) C_b^{B\times B}(z_2,z_4) + \\
    &C_b^{A\times B}(z_1,z_4) C_b^{B\times A}(z_2,z_3)],
\end{split}
\end{equation}
where $f_{\rm sky}$ is the fraction of the sky seen by the survey and $\Delta \ell$ is the width of the bin centered at $\ell$. For intensity mapping tomography, the first term is the product of the autopower of the IM data $C_b^{\rm IM}$ and the galaxy survey $C_b^{g}$, and the second term is the product of cross-powers $C_b^\times$ between the IM data and the galaxy survey. For the \FB\ analysis, the first term dominates the covariance due to the high thermal noise and large (at low $\ell$) Milky Way foreground signal present in $C_b^{\rm IM}(z,z')$. So, as a rough rule of thumb, the $b = b'$ terms of the cross-power covariance are
\begin{equation}\label{eq:cross_power_cov_approx}
\begin{split}
    &\langle \Delta C_b^{\times}(z_1,z_2) \Delta C_b^{\times}(z_3,z_4) \rangle \approx \\ &\frac{1}{f_{\rm sky} \Delta \ell (2\ell + 1)} [ C_b^{\rm IM}(z_1,z_3) C_b^{g}(z_2,z_4)].
\end{split}
\end{equation}

Appendix\,\ref{sec:appendix_sensitivity_forecast} applies this equation to derive an approximate formula for the expected sensitivity that a cross-power survey could achieve on the line intensity. The covariance can be more accurately approximated under the assumption of large sky coverage (\cite{tristram2005xspect}, formula included in Appendix\,\ref{sec:Appendix_coupling_approximation}) or computed via simulated draws from an assumed $C_{\ell}(z,z')$ model and repeated \pcl\ computation of the resulting $C_b(z,z')$. For our \BB\ analysis, we use the approximate covariance of \cite{tristram2005xspect}, as it matches our simulations well.
For the \FF\ and \FB\ analysis, the approximation fails to match simulations, so we instead use a fully simulated covariance. We suspect that the inconsistency between the simulations and the approximate formula is due to the combination of small sky coverage and the steep angular index of the Milky Way foregrounds. Further details on the covariance used for each analysis are included in Section\,\ref{subsec:FB_cov}.

\subsection{Implementing the estimator}
\label{subsec:implementing_estimator}

For a range of binned multipoles indexed by bandpower $b$, the computed $C_{b}(z,z')$ is a rank-3 tensor, indexed by $b$, $z$, and $z'$. To perform a likelihood analysis, we define $\mathbf{x}$ as a flattened vector of all the unique elements of $C_{b}(z,z')$. 
\begin{equation}
\mathbf{x} \equiv vec [C_{b}(z,z')].
\end{equation}
Note that for auto-powers, there are only $N_{b} \times N_z \times (N_z+1)/2$ unique elements, since $C_{b}^{A \times A}(z,z') = C_{b}^{A \times A}(z',z)$. For the cross-power, all elements are unique, and $\mathbf{x}$ has a length of $N_{b} \times N_z^2$. 

The likelihood formed from $C_{b}(z,z')$ is well-studied in CMB literature (see e.g., \cite{Hamimeche:2008ai} for a survey of likelihood forms depending on assumptions). If the amplitudes of the spherical harmonics $a_{\ell m}(z)$ of the maps are drawn from an underlying Gaussian $C_{\ell}(z,z')$ distribution, then the resulting measured vector of anisotropies, $\mathbf{\hat{x}}$, is Wishart-distributed. With our bin size of $\Delta \ell =9$, and for the $\ell$-range with signal sensitivity (see Figure\,\ref{fig:variance_vs_ell}), the number of modes contributing to each bin is high enough that the Wishart distribution is reasonably well-approximated by a Gaussian distribution
\begin{equation}
\begin{split}\label{eq:gaussian_dist}
-2\ln\mathcal{L} = &[ \bf{x}(\Theta) - \hat{\bf{x}} ]^T \bf{\Sigma}(\Theta)^{-1} [\bf{x}(\Theta) - \hat{\bf{x}} ]\\
&+ \ln |\bf{\Sigma}(\Theta)| - 2k\ln(2\pi),
\end{split}
\end{equation}
where $\bf{\Sigma}(\Theta)$ is the bandpower covariance matrix of the flattened data vector and includes binned angular $b{-}b'$ coupling induced by incomplete sky coverage, $\hat{\bf{x}}$ is a flattened vector of the $\hat{C}_{b}(z,z')$ estimate computed from the data, and $\bf{x}(\Theta)$ is a flattened vector of the $C_{b}(z,z')$ model, which depends on the science parameters of interest, $\bf{\Theta}$. 

When analyzing multiple datasets, such as galaxy surveys and intensity maps, the likelihood can be expanded to apply to a data vector that concatenates the auto- and cross-powers of all the datasets. So, for the FIRAS and BOSS data, the data vector $\hat{\bf{x}}$ could contain flattened \FF, \BB, and \FB\ $C_{b}(z,z')$ estimates. If the thermal noise and galaxy shot noise were small, then the high correlation between the galaxy data and \cii\ data, which trace the same matter fluctuations, could be exploited to remove cosmic variance from the measurement of the \cii\ line intensity \citep{McDonald:2008sh, Bull:2014rha, 2016MNRAS.455.3871A, switzer2017tracing, Switzer:2018tel, 2021PhRvD.104h3501O}. For this \FB\ analysis, thermal noise and foregrounds, rather than cosmic variance, are dominant sources of error, so the benefits of such an approach are negligible. However, future intensity mapping experiments may benefit from this approach. For simplicity, we focus on \cii\ constraints from the cross-power, \FB, and we only use \FF\ and \BB\ to validate the cross-power covariance model.

\section{The FIRAS and BOSS data sets}
\label{sec:FIRAS_and_BOSS_data}

\subsection{FIRAS Instrument and Data Set}
\label{sec:FIRAS}
\begin{figure}
  \includegraphics[width=\columnwidth]{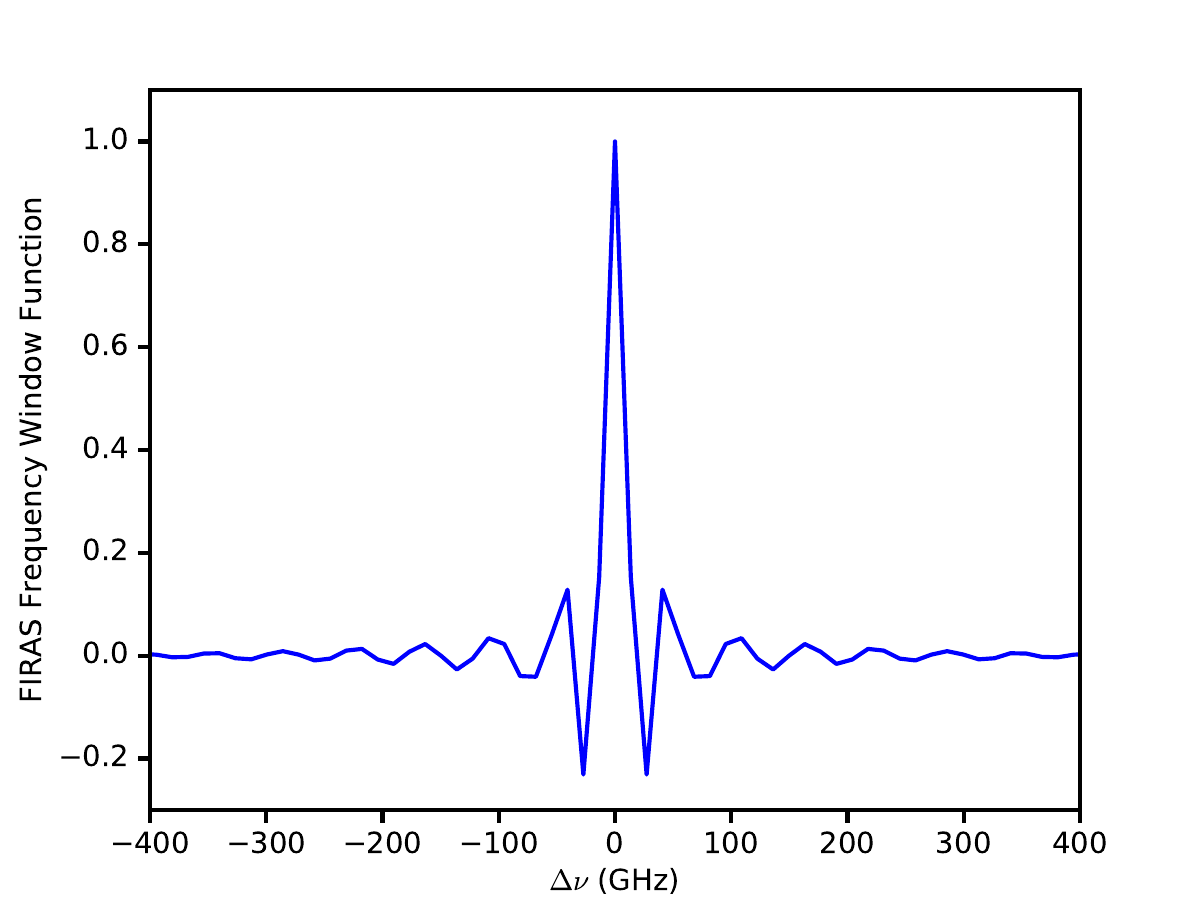}
  \caption{ \label{fig:FIRAS_freq_window_func}Each interferogram measured by FIRAS was multiplied by an apodization function before Fourier transforming. The result is that the FIRAS maps show the true sky spectrum convolved by the Fourier transform of that apodization function. We call this Fourier transformed apodization function the FIRAS frequency response function, $A(\Delta \nu)$, and plot it here. }
\end{figure}
FIRAS is a rapid-scan polarizing Michelson interferometer \citep{mather1993design} on the COBE satellite that mapped the frequency spectrum of the full infrared sky at a coarse angular resolution. The frequency spectrum for each pointing was obtained via an inverse Fourier transform of the interferogram of measured powers over a discrete range of instrument path length differences. The resulting measurements of the sky spectrum are equal to the true spectrum convolved by the inverse Fourier transform of an apodization function \citep{fixsen1994calibration}. Figure\,\ref{fig:FIRAS_freq_window_func} plots this frequency response function, which we shall denote $A(\Delta \nu)$. 

\begin{figure}
  \includegraphics[width=\columnwidth]{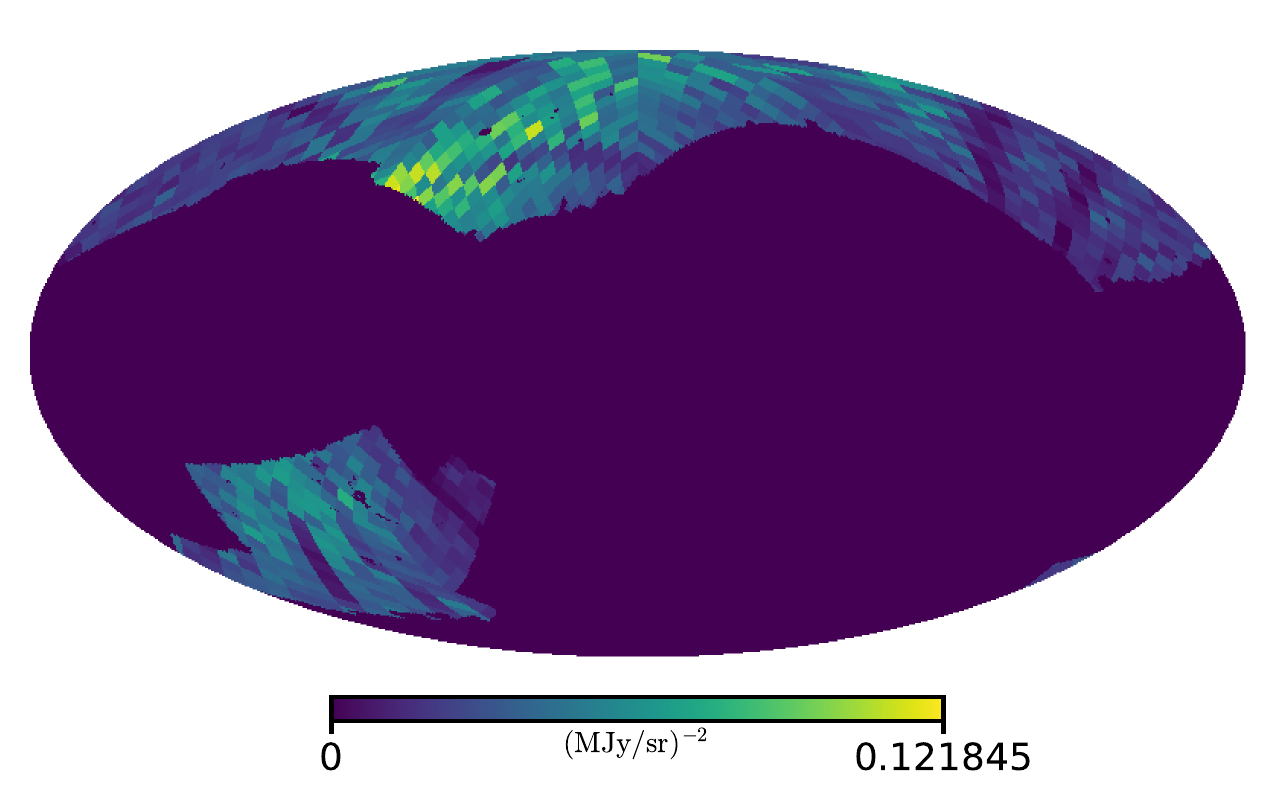}
  \caption{ \label{fig:FIRAS_weights} Mollweide projection of the FIRAS inverse-noise weights for \cii\ emission at $z\sim0.52$. We zero the weights at all regions that do not overlap with the CMASS galaxy survey. Although this plot only shows the weights for the single frequency bin centered at $z\sim0.52$, the angular distribution of the weights is identical for all frequencies. The frequency dependence of the weights can be inferred from the orange curve of Figure\,\ref{fig:Gal_counts_noise}.}
\end{figure}

The published FIRAS maps are binned in the HEALPix \footnote{http://healpix.sourceforge.net} \citep{2005ApJ...622..759G, Zonca2019} format with resolution parameter $N_{\rm side}{=}16$, corresponding to 3072 angular pixels, sufficient to sample the 7-degree beam. In addition to the sky maps, inverse-noise weight maps were produced based on fluctuations of the different interferograms contributing to each pixel. We upgrade the map binning to $N_{\rm side}{=}128$ for analysis. This regridding does not gain any angular information from the FIRAS maps, but it allows finer, more accurate noise weights for the galaxy over-density maps. Figure\,\ref{fig:FIRAS_weights} shows the inverse-noise weights at $N_{\rm side}=128$ for \cii\ emission from $z\sim0.52$. 

\begin{figure}
  \includegraphics[width=\columnwidth]{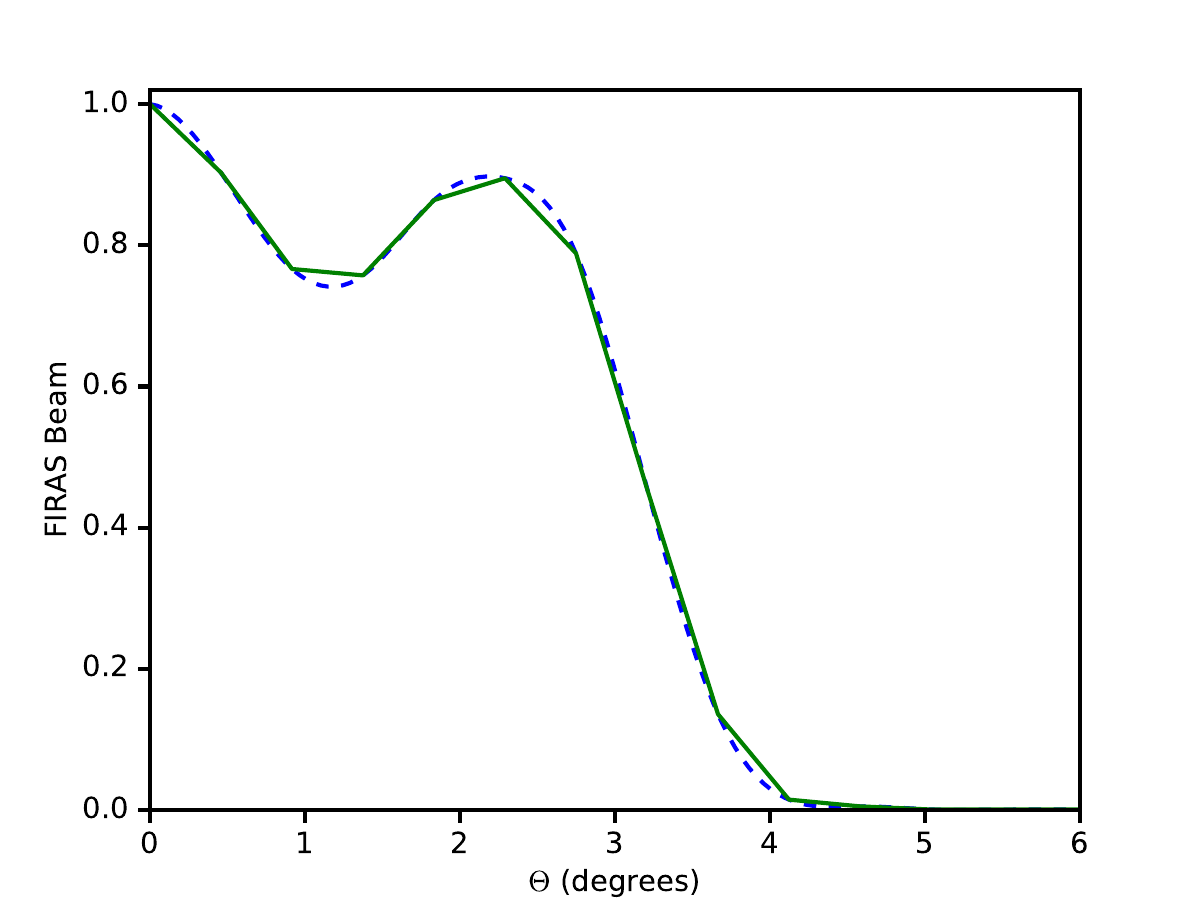}
  \caption{ \label{fig:FIRAS_beam}The measured FIRAS beam (solid green) and our interpolation (dashed blue). The beam was measured at a frequency of $1.5$\,THz via observations of the Moon.}
\end{figure}

\begin{figure}
  \includegraphics[width=\columnwidth]{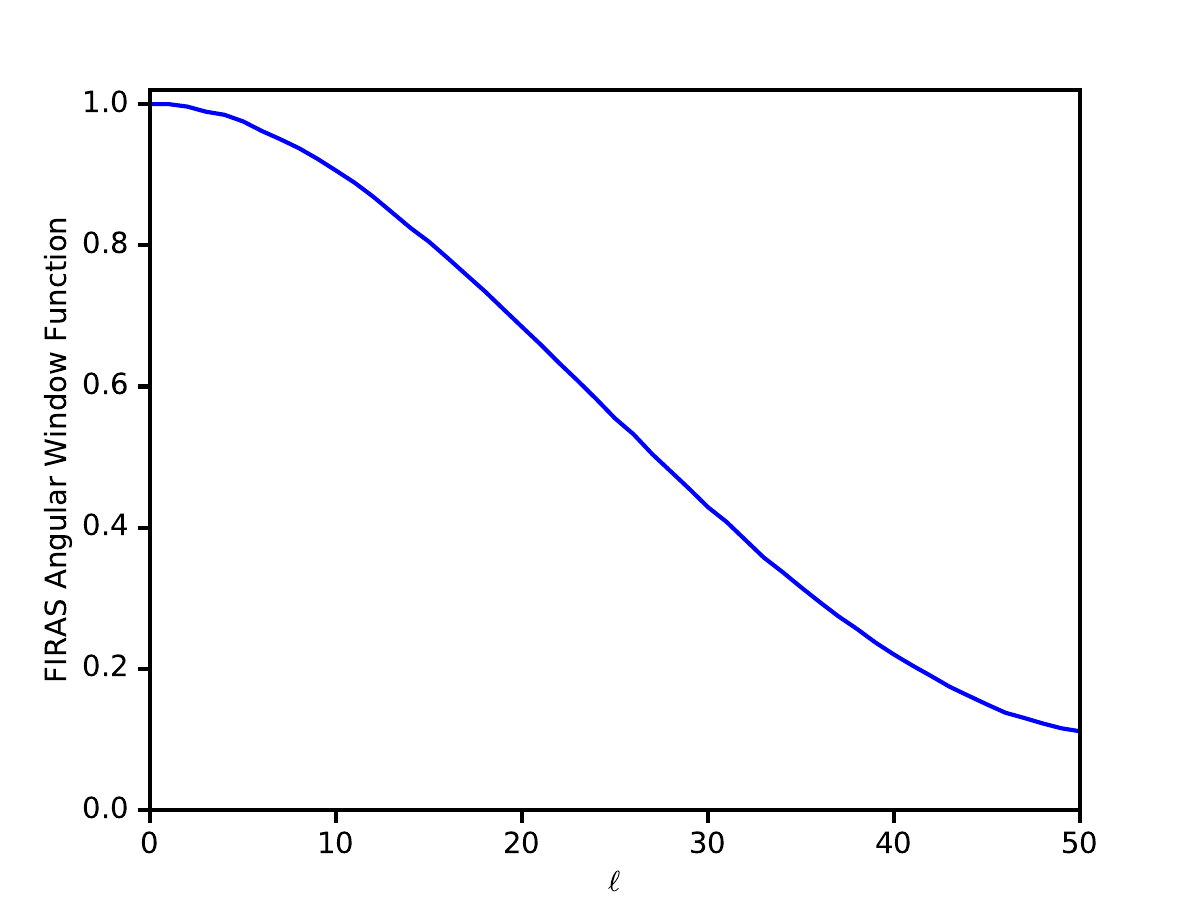}
  \caption{ \label{fig:FIRAS_win}The simulated FIRAS angular window function, $\omega_{FIR}(\ell)$, which accounts for beam, scan, and pixelization effects.}
\end{figure}

The FIRAS beam is formed by a non-imaging parabolic concentrator, which creates a near-tophat beam response with 7-degree FWHM. This beam was measured by in-flight scans of the Moon (Figure\,\ref{fig:FIRAS_beam}). In addition to this intrinsic beam convolution, the finite time required to complete an interferogram combined with the FIRAS scan motion causes the maps to be further smoothed in the ecliptic scan direction by a 2.4-degree tophat function. We account for all beam, scan, and pixelization smoothing effects on the power spectrum via simulation: realizations of a model power spectrum are drawn at $N_{\rm side}=128$, convolved by the FIRAS beam model, convolved by a 2.4-degree tophat function in the ecliptic direction, degraded to $N_{\rm side}=16$, and re-grid onto $N_{\rm side}=128$. We then compute an angular transfer function in $\ell$ by calculating the ratio of the angular power spectrum calculated from these modified maps to the original angular power spectrum. The square root of this transfer function, denoted $\omega_{FIR}(\ell)$ (Figure\,\ref{fig:FIRAS_win}), represents the combined beam, scan, and pixel window function for the FIRAS maps. 

\subsection{BOSS Data Set}
\label{sec:BOSS}
The BOSS survey \citep{dawson2012baryon} extends the Sloan Digital Sky Survey (SDSS). It was designed to measure the Baryon Acoustic Oscillation feature in the galaxy matter power spectrum. Precise spectroscopic redshifts were obtained for approximately 1.5\,million galaxies in the redshift range $0{<}z{<}0.8$, selected to have approximately constant stellar mass. 
Details about the telescope and instruments of SDSS can be found in \cite{1996AJ....111.1748F}, \cite{1998AJ....116.3040G}, \cite{gunn20062}, \cite{2010AJ....139.1628D}, and \cite{smee2013multi}.
\begin{figure}
  \includegraphics[width=\columnwidth]{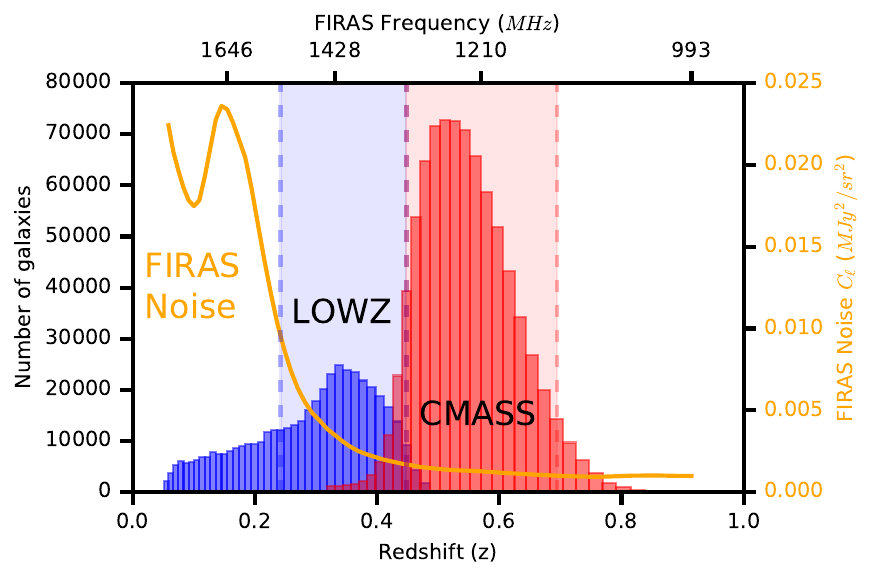}
  \caption{ \label{fig:Gal_counts_noise} Galaxy counts as a function of redshift for the BOSS LOWZ (blue) and CMASS (red) populations. Plotted in yellow is the expected FIRAS noise angular power spectrum, $C_{\ell}$, at $\ell=0$ for the corresponding frequencies of the \cii\ fine-structure line. For most of the angular modes accessible to FIRAS, the noise power spectrum will be higher, growing roughly proportional to the inverse of the square of the FIRAS beam and scan window function. For $z<0.2$, the FIRAS noise grows rapidly. The shaded blue and red rectangles indicate the redshift ranges used for the cross-power analysis of the FIRAS data with the LOWZ and CMASS samples. These regions were selected because both the FIRAS thermal noise and galaxy shot noise are low.}
\end{figure}

\begin{figure}
  \includegraphics[width=\columnwidth]{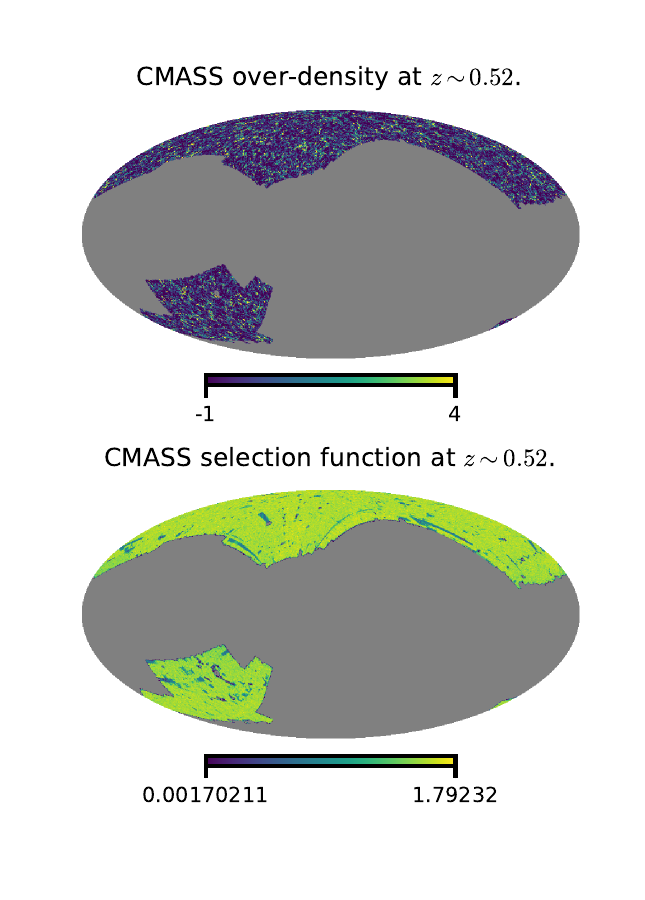}
  \caption{ \label{fig:CMASS_sel} Mollweide projection of the CMASS over-density map and selection function at $z\sim0.52$, near the redshift peak of CMASS, in Galactic coordinates. There are several points in the over-density map well above 4, but we choose to saturate the scale at 4 to show the broad clustering features.} 
\end{figure}

BOSS data release 12 \citep{alam2015eleventh} includes 100 mock unclustered realizations of CMASS and LOWZ galaxies. We bin both the real catalogs and the unclustered mocks onto HealPix maps with $N_{\rm side}=128$, in redshift bins corresponding precisely to the FIRAS frequency bins under the assumption that the FIRAS signal is redshifted \cii\ emission. We construct the CMASS and LOWZ galaxy selection functions, denoted $\bar n(z, \theta)$, by averaging their mock catalogs, assuming a selection function that is separable in angle and redshift. The separability assumption reduces shot noise in the selection function and should be sufficiently accurate for the cross-correlation, limited by FIRAS noise. We define the boundary of the survey by zeroing pixel weights where the selection function is more than $1.3$ standard deviations below the angular average of the sample, which additionally de-weights some regions with lower coverage. The galaxy over-density field is then formed via 
\begin{equation}
\delta^g(z, \theta) = \frac{n(z,\theta)}{\bar n(z, \theta)} - 1,
\end{equation}
where $n(z,\theta)$ denotes the binned galaxy maps and $\bar n(z, \theta)$ denotes the selection function. Figure\,\ref{fig:CMASS_sel} displays the galaxy over-density maps and selection function for a single redshift slice at $z\sim0.52$.

Previous analysis of the BOSS data has found evidence of systematic contamination from stellar populations at low $\ell$ \citep{loureiro2019cosmological}. We also find evidence of systematic low $\ell$ contamination, with spurious $z-z'$ correlations visible at the lowest $\ell$-bin of our analysis ($2\leq \ell \leq 10$), indicative of a common systematic component across redshifts. $\chi^2$ tests of BOSS data compared to our fitted model are high when including this bin. However, they drop to expected values when excluding this bin. We cut this lowest angular bin and the next bin ($11\leq \ell \leq 19$) from our analysis (see Figure\,\ref{fig:variance_vs_ell}). 

\section{Signal Model and Parameter Fit}
\label{sec:signal model}

\begin{figure}
  \includegraphics[scale=0.45]{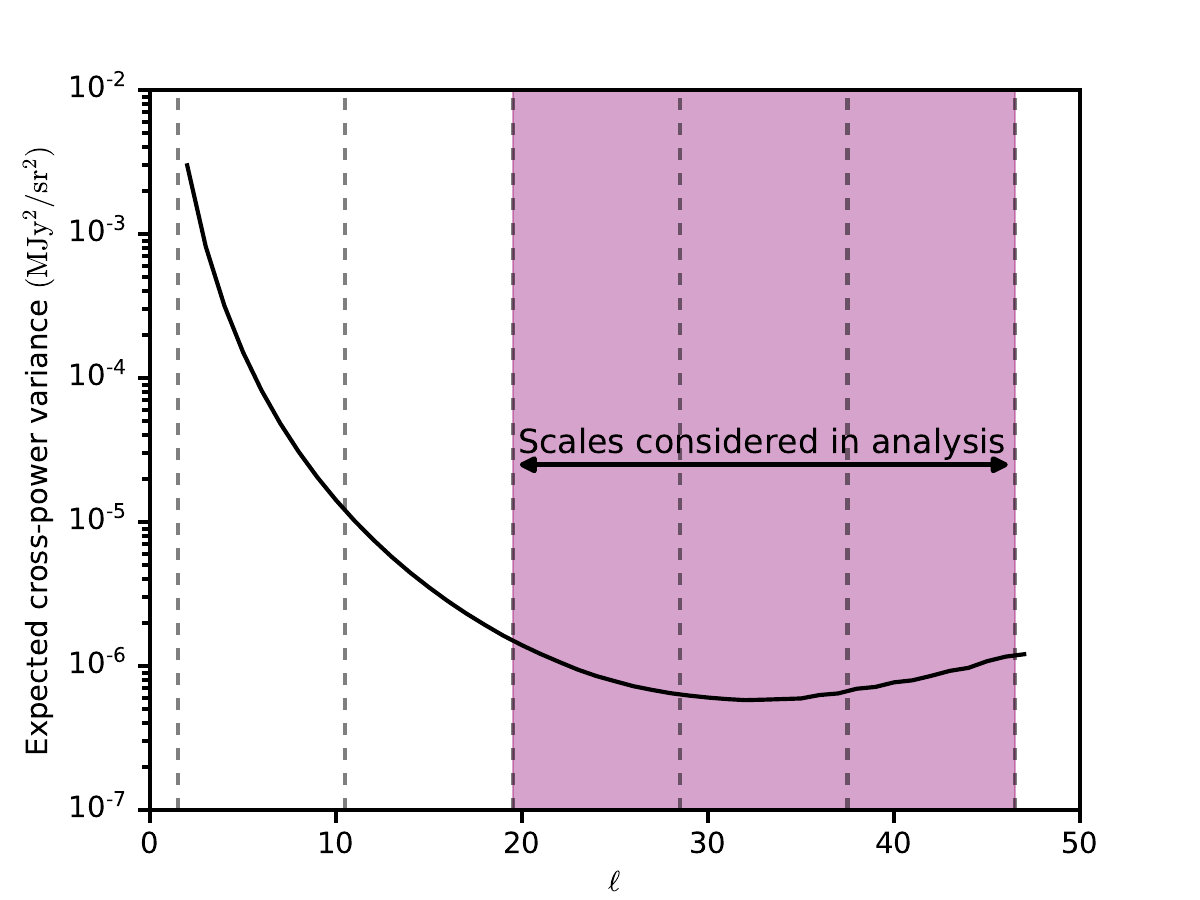}
  \caption{ \label{fig:variance_vs_ell} The approximate expected variance in the cross-power diagonal at $z {\sim} 0.4$, according to the Gaussian error formula $\langle \Delta C^{x}_\ell(z,z') \Delta C^{x}_\ell(z,z') \rangle \approx (f_{\rm sky}(2\ell + 1))^{-1} C^{\rm IM}_\ell(z,z')C_{\ell}^g(z,z')$. Although the magnitude varies somewhat with redshift, the shape is representative of all redshifts studied. The small number of modes and large galactic foregrounds drive high variance at low $\ell$. For $\ell>20$, the foregrounds have mostly subsided, and thermal noise dominates. For $\ell > 30$, the increase in the noise caused by beam and scan convolution starts to overtake the advantage of extra modes at higher $\ell$. Since the FIRAS data was originally mapped at $N_{\rm side}=16$, spherical harmonics below $\ell=3N_{\rm side}=48$ form a complete basis, and no further information can be extracted by considering higher $\ell$. Dashed lines indicate the bounds of the 5 $\ell$-bins, and the shaded regions show the three bins used in the cross-power analysis.}
\end{figure}

The low spatial resolution of the FIRAS maps limits the range of scales to $2 < \ell < 47$, or five bandpowers with $\Delta \ell = 9$. Figure\,\ref{fig:variance_vs_ell} shows the expected variance on the cross-power signal for each $\ell$ and the $\ell$-bins we use in this analysis. Since the first of the five available $\ell$-bins shows signs of stellar contamination in the BOSS sample, we drop it from our analysis. We also eliminate the second $\ell$-bin because, in that bin, the mode mixing caused by partial sky coverage combines with the steep angular index of Milky Way emission to mix the FIRAS auto-power negative. This makes the empirical FIRAS model non-positive definite in that bin and therefore unusable as verification for our cross-power fits. Consequently, we use only the last three bins, as indicated in Figure\,\ref{fig:variance_vs_ell}, which contain most of the sensitivity. In principle, extra information could be obtained at higher $\ell$ by re-making the FIRAS maps from the raw data at higher $N_{\rm side}$, but in practice, the 7-degree beam of the FIRAS instrument diminishes the signal-to-noise ratio at higher $\ell$s (see the upward trend in noise-to-signal ratio for $C_{\ell}^{\times}(z,z')$ at $\ell>35$ in Figure\,\ref{fig:variance_vs_ell}).
To model the errors, we also restrict the \FF\ and \BB\ analysis to these same 3 $\ell$-bins, even though much higher $\ell$ information is available from the BOSS catalog. Similarly, we must restrict the \FF\ analysis to the sky fraction covered by the BOSS galaxy survey to avoid over-estimating the error bars by including the bright Galactic plane, which has no overlap with the BOSS North or South fields.

\subsection{Dark Matter Model}\label{subsec:dark_matter_model}
Both our \BB\ and \FB\ models require the dark matter angular power spectrum of the overdensity field, $C_{\ell}^{\delta}(z,z')$. We calculate this angular power spectrum with the Boltzmann code CLASS \citep{2013JCAP...11..044D, 2014JCAP...01..042D}, using cosmological parameters inferred from the Planck 2015 \citep{Ade:2015xua} temperature and low-$\ell$ polarization maps (TT+LowP). The Halofit routine \citep{Smith:2002dz} provides nonlinear corrections to the power spectrum. 
However, on the several-degree scales of this analysis, the fluctuations are well-described by linear perturbation theory ($k_{\rm max} \sim \frac{\ell_{\rm max}}{\chi(z)_{\rm min}} \sim \frac{50}{880}$ h/Mpc $\sim 0.06$ h/Mpc), and nonlinear corrections are small. 
CLASS computes the angular power spectrum from the 3D power spectrum, $P(k)$, according to the equation
\begin{equation}
C_{\ell}^{A\times B}(z,z') = \frac{2}{\pi} \int k^2 P^{\delta}(k,z{=}0) W_A^{\rm tot}(k,z)W_B^{\rm tot}(k,z') dk,
\end{equation}
where $P^{\delta}(k,z{=}0)$ is the dark matter power spectrum at the current epoch, and, if there are no redshift space distortions (RSDs),
\begin{equation}
W_A(k, z) = b_A \int \phi_z(z'') G(z'',k) j_{\ell}[k \chi (z'')] dz'',
\end{equation}
where $b_A$ is the bias for dark matter tracer $A$, $\phi_z(z'')$ is a tophat redshift selection function that is non-zero only over the range of the redshift slice centered at redshift $z$ and normalized to integrate to 1, $ j_{\ell}$ is a spherical Bessel function of the first kind with parameter $\ell$, $G(z'',k)$ is the growth factor, and $\chi(z'')$ is the radial comoving distance to the shell at redshift $z''$. Linear redshift space distortions (RSDs) can be included \citep{Fisher:1993pz, padmanabhan2007clustering} by replacing $W_A(k, z)$ with $W_A^{\rm tot}(k,z)$, where
\begin{equation}
\begin{split}\label{eq:RSD_boss}
W_A^{\rm tot}(k, z) = &W_A(k, z) + W_A^{\rm RSD}(k,z)\\
W_A^{\rm RSD}(k,z) =&b_A \int  \beta_A(z'') \phi_z(z'')  \times \\
\biggl [ &\frac{2\ell^2+2\ell -1}{(2\ell+3)(2\ell-1)}j_{\ell}(k \chi (z''))\\ 
&- \frac{\ell(\ell-1)}{(2\ell-1)(2\ell+1)}j_{\ell-2}(k \chi(z''))\\ 
&- \frac{(\ell+1)(\ell+2)}{(2\ell+1)(2\ell+3)}j_{\ell+2 }(k \chi(z'')) \biggr ] dz'',\\
\end{split}
\end{equation}
\\
where $\beta_A(z) = f(z)/b_A$, with $f(z)$ being the logarithmic growth rate of linear perturbations in the matter power spectrum.  The $C_{\ell}^{A\times B}(z,z')$ that results from using $W_A^{\rm tot}(k, z)$ and $W_B^{\rm tot}(k, z)$ contains cosmological terms proportional to $b_A b_B$, proportional to $(b_A {+} b_B)/2$, and independent of both $b_A$ and $b_B$. We label these terms $C_{\ell}^{(2)}(z,z')$, $C_{\ell}^{(1)}(z,z')$, and $C_{\ell}^{(0)}(z,z')$ respectively. They are calculated from CLASS via linear combinations of $C_{\ell}(z,z')$ computations without RSD and bias 1, with RSD and bias 1, and with RSD and bias 0. With this formalism, the cross-power spectrum of two biased matter tracers is given by 
\begin{equation}
C_{\ell}^{A\times B}(z,z') = b_A b_B C_{\ell}^{(2)}(z,z') + \frac{b_A {+} b_B}{2}C_{\ell}^{(1)}(z,z') + C_{\ell}^{(0)}(z,z').
\end{equation}
The bias dependence of these terms is reminiscent of the Kaiser correction in power spectrum space. Indeed, these equations can be derived by including the Kaiser enhancement term in a plane-wave expansion of the power spectrum and integrating along the line-of-sight \citep{padmanabhan2007clustering}.

We do not model finger-of-God effects or redshift smearing due to spectroscopic survey errors since our redshift bin size of $\Delta z \sim 0.02$ is more than 10 times larger than the 400 km/s satellite galaxy velocity dispersion fit and 30 km/s spectroscopic error fit found by \cite{Guo:2014iga}'s analysis of CMASS galaxies. Analyses with sufficient redshift resolution to resolve the finger-of-God effect can include it in the signal model by smoothing the radial window function in the $C_{\ell}(z,z')$ calculation \citep{2019MNRAS.485..326L, 2020PhRvD.102h3521G}.

\subsection{\texorpdfstring{\FB}{FIRASxBOSS}}\label{subsec:cross-power}
The cross-correlation signal model consists solely of correlated continuum emission (dust) and \cii\ emission from the BOSS galaxies. Because they are uncorrelated with the cosmological overdensity field, the thermal noise and foregrounds contribute zero average cross-power and thus will not factor into the mean signal model (although the variance caused by their spurious correlation with the galaxy survey will be included through the FIRAS auto-power in the covariance).  

In Appendix\,\ref{sec:Appendix_A}, we derive the functional form of our cross-power model. The \cii\ part of the model is
\begin{equation}\label{eq:cii_cl_full_orig}
\begin{split}
&C_{\ell}^{\rm \cii \times g}(z,z') = \\
&I_{\rm \cii}(z) \cdot \Big[ b_gb_{\rm \cii}C^{(2)}_{\ell}(z,z') + \\ &\frac{b_g + b_{\rm \cii}}{2}C^{(1)}_{\ell}(z,z') + C^{(0)}_{\ell}(z,z') \Big]. 
\end{split}
\end{equation}
The redshift dependence of $I_{\rm \cii}(z)$ is shown in Equation\,\ref{eq:I_CII}. An intensity mapping experiment with sufficient sensitivity can fit the parameters that control the redshift evolution of $I_{\rm \cii}(z)$. However, due to the high noise and large beam of FIRAS, we fix all of those evolution parameters with reasonable values from \cite{pullen2018search} (see details in Appendix\,\ref{sec:Appendix_A}). Those values lead to a modest evolution in which the brightness increases ${\approx}20\%$ toward higher redshift over each of the CMASS and LOWZ redshift ranges. Our MCMC analysis assumes this redshift shape and fits for the overall amplitude of $b_{\rm \cii}I_{\rm \cii}(z{=}z_{\rm center})$ at the central redshift of each region (CMASS and LOWZ, respectively).

The CIB portion of the cross-power model is
\begin{equation}\label{eq:cib_cl_full_orig}
\begin{split}
&C_{\ell}^{c \times g}(z,z') = \\
& \sum_{z''} \frac{dI_{\rm CIB}(\nu_{\rm \cii}^z,z'')}{dz''}\Delta z'' \times \left[ b_gb_{\rm \cii}C^{(2)}_{\ell}(z'',z') + \right.\\
&\left. \frac{b_g + b_{\rm \cii}}{2}C^{(1)}_{\ell}(z'',z') + C^{(0)}_{\ell}(z'',z') \right],
\end{split}
\end{equation}
where $\frac{dI_{\rm CIB}(\nu_{\rm \cii}^z,z'')}{dz''}\Delta z''$ is the intensity of the CIB that is emitted from sources in a redshift bin of size $\Delta z''$, centered at redshift $z''$, and measured at a frequency of $\nu_{\rm \cii}/(1+z)$.
The sum over $z''$ in Equation\,\ref{eq:cib_cl_full_orig} should, in principle, be carried out over all redshifts, even those outside of the galaxy survey. In practice, for the redshift bins and $\ell$ bins considered in this analysis, the bracketed $C_{\ell}(z'',z')$ kernel is dominated by the $z''= z'$ term, with the neighboring off-diagonal terms around 20\% of the diagonal term, and the rest of the terms are negligible. As with our \cii\ analysis, because of the limited sensitivity of the FIRAS data, we do not attempt to constrain the redshift evolution of the CIB brightness or the spectral shape of the CIB emission. Instead, these are fixed by the assumed values of $\beta = 1.5$ \citep{Planck2014}, $T_d=26$\,K \citep{Serra:2014pva}, $\Phi(z) = (1+z)^{2.3}$ \citep{pullen2018search}, $\log_{10}(M_{\rm eff}/M_{\odot}) = 12.6$ \citep{Planck2014, serra2016dissecting}, and $\sigma^2_{L/M}=0.5$ \citep{10.1111/j.1365-2966.2012.20510.x, Planck2014, serra2016dissecting}. Because of this assumed spectral and redshift evolution, the parameter we constrain is $\frac{dI_{\rm CIB}(\nu_{\rm \cii}^z = \nu_{\rm center} ,z'' = z_{\rm center})}{dz''}$, where $z_{\rm center}$ is the central redshift of the analysis region, and $\nu_{\rm center}$ is the corresponding central frequency of the analysis region.

Finally, our signal model must account for the FIRAS data being convolved by the FIRAS frequency response function, $A(\nu)$. To do this, we convert the frequency response function $A(\nu)$ to a function of redshift, $A(z)$, and convolve $C_{\ell}^{\times}(z,z')$ by $A(z)$, resulting in the final signal model
\begin{equation}
C_{\ell}^{\times}(z,z') = A(z'') \circledast \left[ C_{\ell}^{\rm \cii \times g}(z'',z') + C_{\ell}^{c \times g}(z'',z') \right].
\end{equation}

In our MCMC analysis, we fix the value of the galaxy bias, $b_g$, to the best-fit value from our \BB\ analysis (Section\,\ref{subsec:BOSS_auto_model}) and also fix the \cii\ and CIB bias to be identical to the galaxy bias ($b_{\rm \cii}=b_g$). This fixing of the \cii/CIB bias has almost no effect on our fits since only the small RSD terms ($C^1_{\ell}(z,z')$ and $C^0_{\ell}(z,z')$) can break the degeneracy between $b_{\rm \cii}$ and $I_{\rm \cii}$, and \FB\ lacks the precision to break that degeneracy. The next section, \ref{subsec:FB_cov} develops the covariance model employed by the MCMC.

\begin{figure}
  \includegraphics[width=\columnwidth]{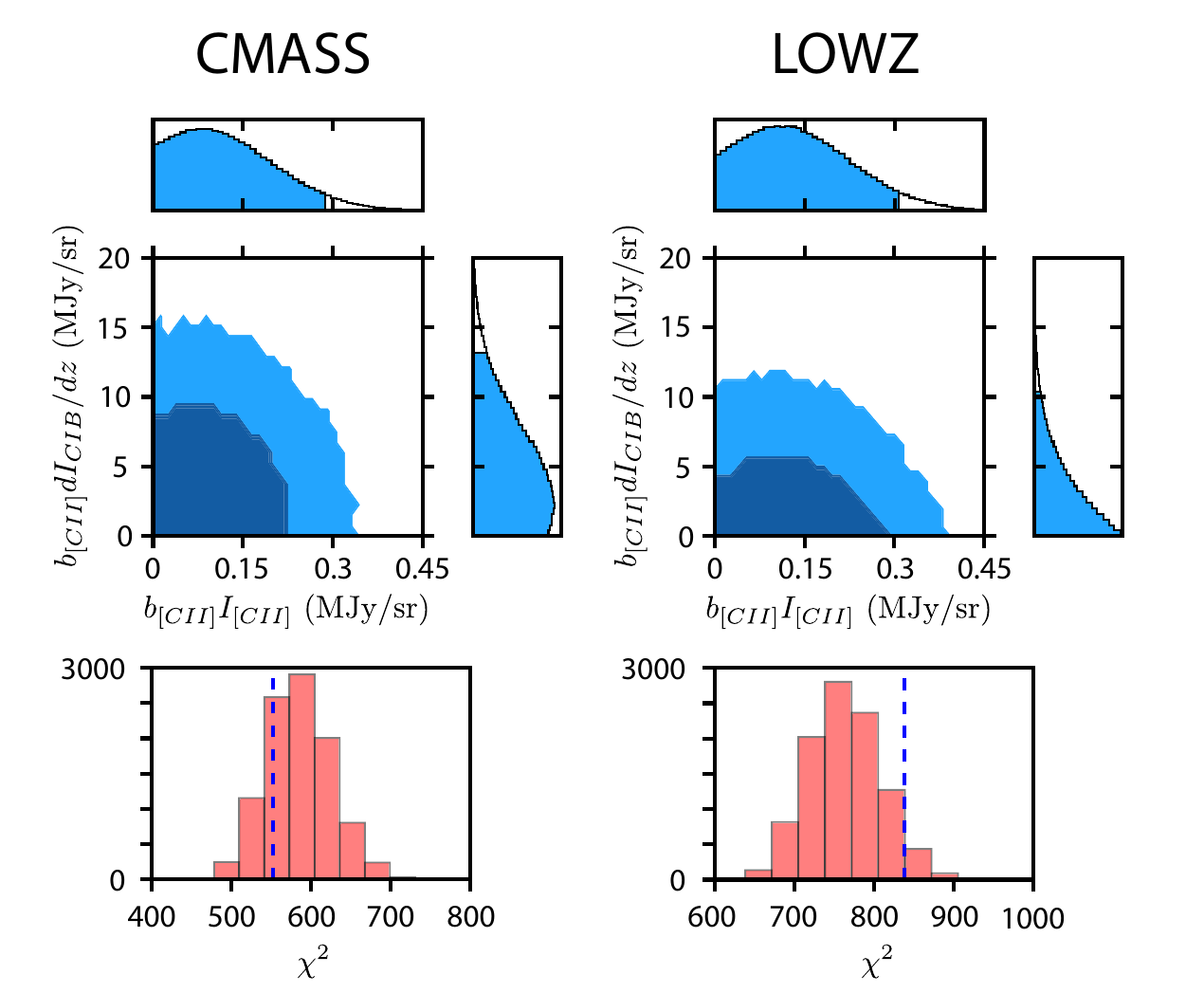}
  \caption{Cross-power fits for the line and continuum amplitude from \FB, showing the CMASS analysis on the left and the LOWZ analysis on the right. The top row shows the MCMC contours of our fit to the data. Dark blue and light blue regions represent the 68 and 95 percent contours, respectively.  The MCMC analysis uses a simple flat prior that restricts both $b_{\rm \cii}I_{\rm \cii}$ and $b_{\rm \cii}dI_{\rm CIB}/dz$ to positive values. This prior has a minimal effect on our best-fit $b_{\rm \cii}I_{\rm \cii}$ values, but it serves to prevent nonphysical negative values from counting towards our upper limit constraints. The bottom row shows red histograms of the distribution of best fit $\chi^2$ for 10,000 simulations in which we draw $a_{\ell m}$ amplitudes from Gaussian distributions for the cosmological signal, galaxy shot noise, and FIRAS auto-power (for more details of the simulation, refer to the parametric simulations described in Appendix\,\ref{sec:Appendix_cov_sims}). The $\chi^2$ of the maximum likelihood fits to the data are plotted as vertical dashed blue lines.}
  \label{fig:cross_power_fits}
\end{figure}

Figure\,\ref{fig:cross_power_fits} shows the MCMC contours for our parameter fits to the \FB\ data with the CMASS and LOWZ galaxies. Also shown, on the bottom row, are the $\chi^2$ of the CMASS and LOWZ maximum likelihood fits, compared to a simulated distribution of best-fit $\chi^2$ values. For the MCMC analysis, we apply a simple flat prior that restricts both $b_{\rm \cii}I_{\rm \cii}$ and $b_{\rm \cii}dI_{\rm CIB}/dz$ to positive values. This prior has almost no effect on our best-fit $b_{\rm \cii}I_{\rm \cii}$ values, but since we do not have enough sensitivity for detection, it prevents unphysical negative values from counting towards our quoted upper limit constraints. We find, at 95 percent confidence, that $b_{\rm \cii}I_{\rm \cii}<0.31$\, MJy/sr at $z\sim0.35$ and $b_{\rm \cii}I_{\rm \cii}<0.28$\, MJy/sr at $z\sim0.57$. 

\subsection{Modeling the \texorpdfstring{\FB}{FIRASxBOSS} Covariance}
\label{subsec:FB_cov}
The required pieces for the covariance used in the MCMC parameter estimation are models for:  1) the \cii\ and CIB signal associated with cosmological clustering, 2) the BOSS signal associated with cosmological clustering, plus shot noise, and 3) the FIRAS thermal noise and foregrounds. In the following three subsections, we describe each of these models in turn. Appendix\,\ref{sec:Appendix_B} describes our method of simulating the covariance from these three models.

\subsubsection{\texorpdfstring{\cii}{[CII]} and CIB signal}
We simulate the \cii\ and CIB variance models through map-space simulations that include FIRAS instrumental effects. The \cii\ and CIB signals are painted onto maps drawn from the cosmological clustering signal with linear bias.  
For the \cii\ signal, this is accomplished by multiplying the drawn maps by $I_{\rm \cii}(z)$. For the CIB signal, the maps are matrix-multiplied by $\frac{dI_{\rm CIB}(\nu_{\rm \cii}^z,z'')}{dz''}\Delta z''$, in a map-space analogy to Equation\,\ref{eq:cib_cl_full_orig}. These maps are then convolved by the FIRAS redshift response function, $A(z)$. 

The magnitude of the portion of the covariance that comes from this \cii\ and CIB signal is a function of the \cii\ and CIB amplitudes. In order to account for this, our covariance is constructed from a linear combination of four separate simulations (accounting for each cross-term). Appendix\,\ref{sec:Appendix_B} describes this process.

\subsubsection{Clustering signal and shot noise from \texorpdfstring{\BB}{BOSSxBOSS}}\label{subsec:BOSS_auto_model}
\begin{figure}
  \includegraphics[width=\linewidth]{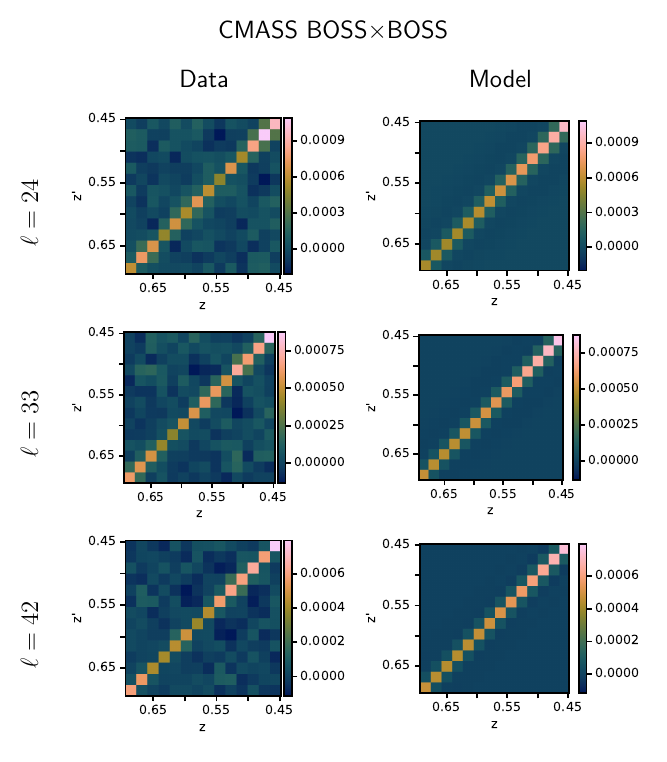}
  \caption{\label{fig:CMASS_gal_model_vs_data} The \BB\ binned angular power spectrum, $C_b(z,z')$, for the CMASS galaxies. The left and right columns show the data and the best-fit model, respectively. The rows show the three angular bins used in this analysis, each of size $\Delta \ell=9$ and centered at $\ell=24$, 33, and 42. At the resolution of FIRAS, most of the cosmological clustering signal occurs on the diagonal, where $z=z'$, though there is also a small correlation between neighboring redshift bins, visible just off of the diagonal. The variation in amplitude along the diagonal is due to the redshift evolution of the growth factor, the change in physical scales being probed as a function of redshift, and changes in the CMASS galaxy density.}
\end{figure}

We model the variance in the BOSS survey as a tracer of the dark matter with constant bias and linear RSD plus shot noise, or
\begin{equation}
\begin{split}\label{eq:boss_full_model}
C_{\ell}^g(z,z') &= b_g^2C_{\ell}^{(2)}(z,z') + b_gC_{\ell}^{(1)}(z,z')\\
&+ C_{\ell}^{(0)}(z,z') + A_{SN}\frac{\Omega_{\rm pixel}}{\bar n(z)}\delta(z,z'),
\end{split}
\end{equation}
where $\bar n(z)$ is the average number of BOSS galaxies per pixel in each redshift slice, and $\Omega_{\rm pixel}$ is the angular size of a pixel in steradians. While the complete \BB\ power spectrum requires additional modeling to describe all scales \citep{loureiro2019cosmological}, this simple model with free parameters for a constant bias and shot-noise amplitude is sufficient to describe the BOSS variance relevant to the angular scales analyzed in \FB. 
Due to the large number of galaxies in the CMASS and LOWZ samples and the limited $\ell$-range of our analysis, the shot noise is considerably smaller than the clustering signal, so $A_{SN}$ is weakly constrained.
The bias parameter is fit to a value of 1.81 and 1.82 for the LOWZ and CMASS samples, consistent with previous work \citep{salazar2017clustering}. We find reasonable $\chi^2$ values for both the CMASS ($\chi^2$ per degree of freedom of 1.02, PTE of 0.37) and LOWZ ($\chi^2$ per degree of freedom of 1.08, PTE of 0.12) fits. Figure\,\ref{fig:CMASS_gal_model_vs_data} shows the measured $C_b (z,z')$ from CMASS versus the best-fit model for the three $\ell$-bins we consider.

The covariance we use for the \BB\ parameter estimation is computed using the approximate formula of \cite{tristram2005xspect} described in Appendix\,\ref{sec:Appendix_coupling_approximation}.
Since the covariance is a function of the model, an MCMC analysis must, in principle, recalculate the covariance for each different estimate of the underlying model parameters. 
For \BB, we instead employ an iterative approach. We make a best guess of the parameters, compute a covariance for that guess, find a new maximum likelihood solution using that covariance, and then repeat the procedure. After several iterations, this procedure has converged, and the parameters assumed in the model for the covariance equal the parameters at the maximum likelihood peak of our MCMC fit to within a fractional tolerance of $3 \times 10^{-3}$. 

\subsubsection{Thermal noise and foregrounds from \texorpdfstring{\FF}{FIRASxFIRAS}}
\label{subsec:FIRAS_auto}

We model the FIRAS thermal noise from the inverse noise variance weights provided by the FIRAS collaboration \citep{FIRASexplanatory}. Let $W(\theta, z)$ represent the FIRAS inverse noise variance maps at $N_{\rm side}{=}128$. Thermal noise contributes constant variance in $\ell$-space, given by $N(z){=}\Omega_{128} \frac{\langle W(\theta,z)\rangle_{\theta}}{\langle W^2(\theta,z)\rangle_{\theta}}$, where $\Omega_{128}$ is the pixel size in steradians, and the angular average is taken only over pixels that overlap the BOSS galaxy survey. 
We then model the FIRAS thermal noise as
\begin{equation}\label{eq:thermal_noise_model}
N(\ell, z, z') \equiv N(z)^{1/2}N(z')^{1/2}A\left(|\nu(z) - \nu(z')|\right),
\end{equation}
where $A\left(|\nu(z) - \nu(z')|\right)$ accounts for the convolution of the FIRAS spectrum by the Fourier Transform Spectrometer's frequency response function. 
The measured frequency correlations of the thermal noise agree with the covariance model of the FIRAS collaboration \citep{FIRASexplanatory}. For this thermal noise component, we compute the expected binned angular power spectrum (Equation\,\ref{eq:model_mixing_binning}) by applying the $N_{\rm side}=16$ pixel window function to $N(\ell, z, z')$, binning into bandpowers, and then unmixing with the binned mixing matrix that uses the full simulated beam, scan, and pixel window function. The result is that, after the binning and unmixing operator is applied, the initially flat angular spectrum of the noise now rises roughly as the inverse square of the FIRAS beam and scan window function. 

\begin{figure}
    \includegraphics[width=0.45\textwidth]{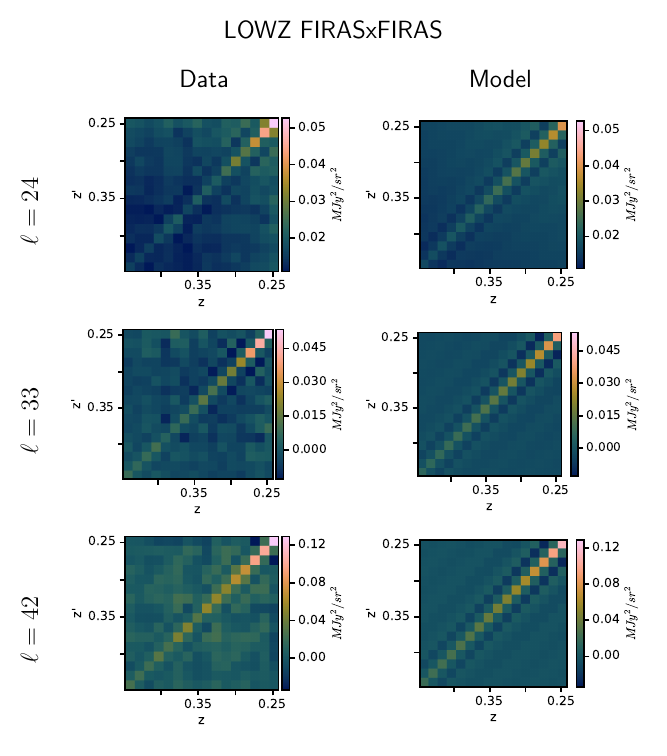}
    \caption{ \label{fig:FIRAS_auto_LOWZ} The \FF\ binned angular power spectrum, $C_{b}(z,z')$, in the LOWZ redshift range for three angular bins of width $\Delta \ell = 9$, centered at  $\ell=24$, 33, and 42. The left column shows the data, and the right column shows the best-fit model, of the form of Equation\,\ref{eq:FIRAS_auto_model}. The structure on the diagonal of the plots is due to thermal noise, with small just-off-diagonal correlations due to the FIRAS frequency window function, $A(\nu)$. The thermal noise increases with higher $\ell$, roughly as the inverse of the FIRAS beam and scan window function squared. The foregrounds are visible in the $\ell=24$ bin as a roughly constant offset to all $z, z'$ combinations. The foreground amplitude is comparable to the thermal noise at $\ell=24$, but it drops at higher $\ell$ with a power-law index of $\gamma \approx -2.3$. The foregrounds are negligible compared to the thermal noise at $\ell=33$ and $\ell=42$. }
\end{figure}

\begin{figure}
    \includegraphics[width=0.45\textwidth]{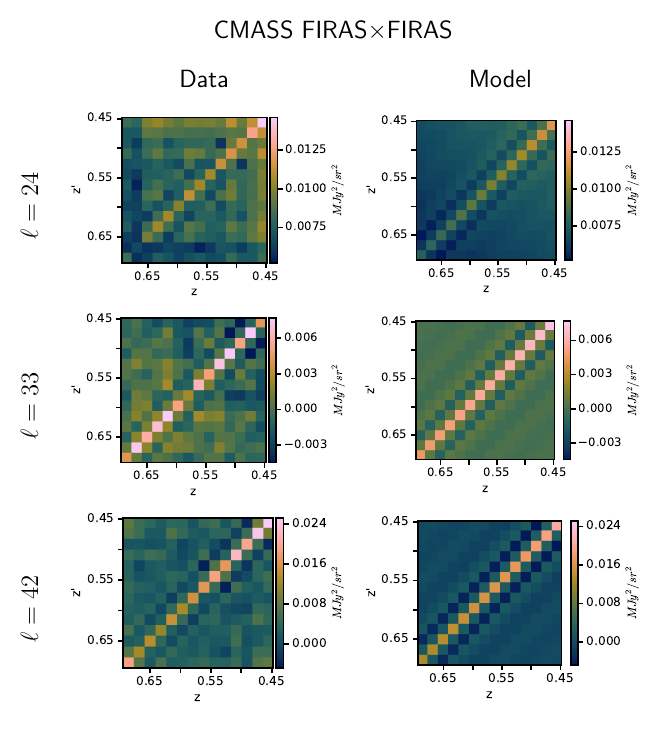}
    \caption{ \label{fig:FIRAS_auto_CMASS} The \FF\ binned angular power spectrum, $C_{b}(z,z')$,  in the CMASS redshift range 
    with layout and general properties similar to the LOWZ region described in Figure\,\ref{fig:FIRAS_auto_LOWZ}.
    The vertical and horizontal lines in the $\ell=24$ data at $z\sim 0.67$ are due to a spurious correlation between thermal noise and Galactic foregrounds. 
    }
\end{figure}

\begin{figure*}
    \includegraphics[width=\textwidth]{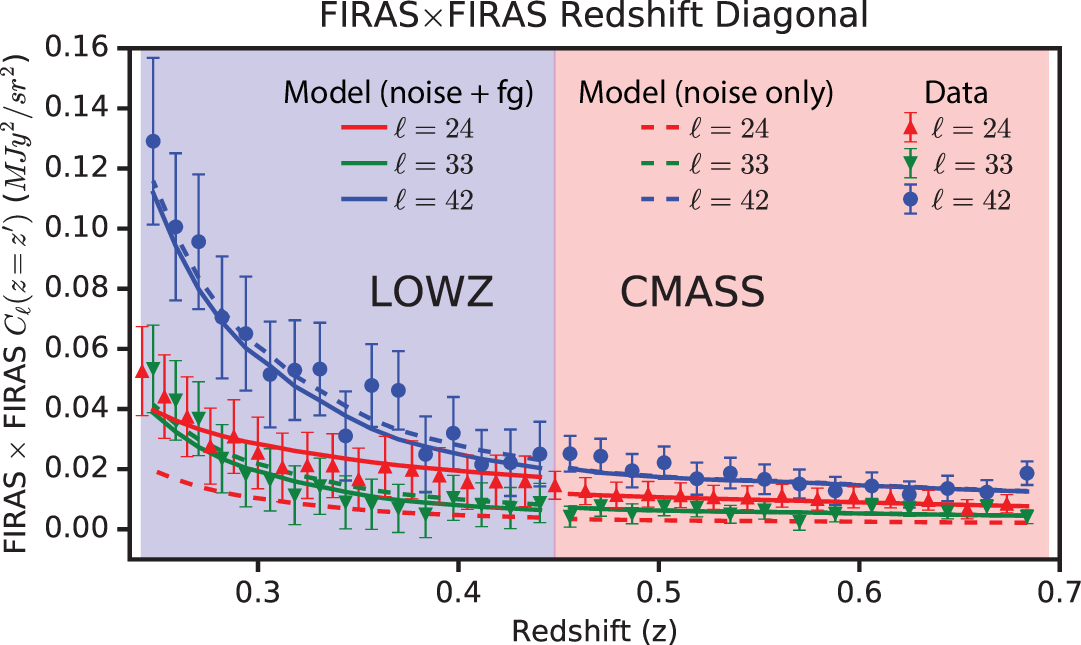}
    \caption{ \label{fig:FIRAS_auto_diag} The redshift diagonal of the \FF\ binned angular power spectrum, $C_b(z=z')$, for both the CMASS and LOWZ redshift regions. The three angular bins, centered at $\ell=24$, $\ell=33$, and $\ell=42$ are plotted in red, green, and blue, respectively. The best-fit models are plotted as solid lines, and the data are plotted as triangles or circles, with error bars computed from the best-fit model. The $\ell=24$ data are artificially shifted backward by half a redshift bin for visual clarity. Dashed lines show the thermal noise portion of the best-fit model only. From the dashed lines, the effect of the FIRAS beam and scan convolution can be seen, as the thermal noise increases from low to high $\ell$. Only the $\ell=24$ bin has a significant foreground contribution.}
\end{figure*}

We model the Milky Way foreground angular power spectrum as a simple power-law with a free angular index $\gamma$ and amplitude $A_{\rm MW}^2$ that we fit to the FIRAS auto-power spectrum through the form
\begin{equation}
C^{\rm MW}_{\ell}(z,z') =   A_{\rm MW}^2 \ell^{-\gamma}D(z,T_d)D(z',T_d).
\end{equation}
The spectrum of the Galactic emission, $D(z,T_d)$, is modeled as semi-thermal dust emission, given by
\begin{equation}
    D(z,T_d) \propto \nu^{\beta} B_{\nu}(T_d), 
\end{equation}
where $\nu$ is converted to redshift assuming the \cii\ line, $\beta = 1.5$ \citep{2014A&A...566A..55P}, $B_{\nu}$ is the Planck function, and the dust temperature $T_d$ is a free parameter. In principle, the Milky Way emission is also convolved in the frequency direction by the frequency response function of the FIRAS spectrometer, but the effect is negligible for smooth spectral emission, so we do not include it. The full model we fit to the data is
\begin{equation}
\label{eq:FIRAS_auto_model}
C^{\rm IM}_\ell(z,z') = A_{\rm MW}^2 \ell^{-\gamma}D(z,T_d)D(z',T_d) + A_N N(\ell, z, z').
\end{equation}
There are four free parameters: the Milky Way amplitude $A_{\rm MW}$ at $\ell\sim1$ (units of MJy/sr), the dust temperature $T_d$ (units of K), the unitless angular power-law index $\gamma$, and a unitless factor $A_N$ multiplying the expected noise signal ($A_N$ is expected to be near 1).
Figures \ref{fig:FIRAS_auto_LOWZ} and \ref{fig:FIRAS_auto_CMASS} show color plots of the best-fit models and the data for LOWZ and FIRAS respectively, over the full redshift and angular range of our analysis. Figure\,\ref{fig:FIRAS_auto_diag} shows the redshift diagonal of the data and best-fit models for both the LOWZ and CMASS redshift ranges, along with error bars estimated from Monte Carlo simulations drawn from the best-fit model, fully including the effects of FIRAS beam convolution, ecliptic scan convolution, pixelization, and partial sky coverage.

The covariance we use to fit \FF\ to our parametric model is simulated. We draw  $a_{\ell m}$ amplitudes from a Gaussian distribution whose variance is given by our model (Equation\,\ref{eq:FIRAS_auto_model}). We use this to produce 10,000 full-sky maps, to which we apply our FIRAS window function. We then use the FIRAS inverse-noise weights and window function to compute a simulated observed binned partial-sky $C_b(z,z')$ for each of these 10,000 draws. The covariance computed from these simulated $C_b(z,z')$ amplitudes accounts for the effects of beam convolution, ecliptic scan smearing, and partial sky coverage.
Since the covariance is a function of the assumed parameters for the model, it should, in principle, be recalculated for each different estimate of the underlying model parameters. For a simulated covariance, this procedure is computationally expensive. As with \BB, we instead employ an iterative approach for \FF. We make a best guess of the parameters, simulate a covariance for that guess, find a new maximum likelihood solution using that covariance, and then repeat the procedure. 

We repeat the above iteration until the $\chi^2$ per d.o.f. converges to ${<}1\%$. Figures \ref{fig:FIRAS_auto_LOWZ}, \ref{fig:FIRAS_auto_CMASS}, and \ref{fig:FIRAS_auto_diag} show the parametric model compared to the measured \FF\ power spectrum. The fit converges to a $1\%$ constraint on the thermal noise amplitude centered on 0.97 for LOWZ and 0.90 for CMASS. The three foreground parameters are correlated and weakly constrained but yield dust temperatures consistent with the $15-25$\,K measured over the region of the sky we observe \citep{PhysRevD.95.103517}. Milky Way emission contributes approximately half of the variance to the $\ell=24$ bandpower and is negligible at the other two bandpowers. Because there are relatively few spatial modes in the BOSS regions at $\ell=24$, and because the BOSS region is relatively clear of Milky Way emission, the constraint on the Milky Way contribution to $\ell=24$ is uncertain to $30\%$, but this uncertainty has a small impact on the final bounds on line brightness from \FB. Since the foregrounds account for half of the redshift-diagonal variance at $\ell=24$, a $30\%$ increase would cause a $15\%$ increase in total $\ell=24$ bin variance. The $\ell=24$ bin makes up roughly half of the total \cii\ signal-to-noise ratio, so this could at most increase the \cii\ variance by $7.5\%$. In amplitude, a $3.75\%$ increase in the final CII constraint is 0.01 MJy/sr. In reality, this is conservative because the foregrounds have long frequency correlations that the \cii\ signal does not have, so most of the impact would instead be on the CIB constraint.

The best-fit parametric models have low $\chi^2$ per d.o.f. (0.74 for LOWZ and 0.86 for CMASS), indicating that the error model may be overestimated. An alternative to the parametric model approach is instead to use the measured auto-power spectrum of the FIRAS data, $\hat{C}_b^{\rm IM}$, as the model of variance in the FIRAS data cube. The attraction of this approach is that it captures features that could be missing in any parametric model fit. Unfortunately, because the FIRAS auto-power signal also contains the \cii\ and CIB signal, this approach introduces cosmic bias wherein high-scattering modes with more \cii\ and CIB signal are artificially down-weighted in the cross-power analysis. This could bias both the measured \cii\ signal and its error bars low, an effect we have measured in simulations.
Although the measured FIRAS auto-power cannot be used in a covariance model acting on the actual cross-power data, we use it on simulated cross-power signal to verify that our parametric model produces similar error bars on the \FB\ parameters. Appendix\,\ref{sec:Appendix_cov_sims} shows the results of two sets of simulations, one with a covariance that uses the measured FIRAS auto-power and one with a covariance that uses our best-fit parametric model. The two models yield nearly identical error contours on the cross-power parameters.
 
We note that the complete FIRAS covariance has additional terms with complex structure, described in \cite{FIRASexplanatory}, which includes correlations jointly between frequencies $\nu, \nu'$ and position vectors $\hat n, \hat n'$ on the sky. 
In the frequency range studied here and the survey region outside the Galactic plane, this covariance is dominated by thermal detector noise, which is diagonal in $\hat n, \hat n'$, and by the correlated structure across frequencies produced by Milky Way emission, at low multipoles. In measurements with future instruments that achieve high significance, the absolute calibration error must also be included.
The FIRAS auto-power also contains the \cii\, signal and the continuum CIB, but at a low level that negligibly effects our parametric thermal noise and foregrounds fits. In addition, there are several prominent Galactic spectral lines \citep{fixsen1999cobe}. The only line that falls in the frequencies of our analysis is the 205.3 $\mu {\rm m}$ NII line. Although this line is visible in the full-sky FIRAS auto-power spectrum, it is not detectable when we restrict the data to the BOSS survey regions, which are out of the Galactic plane, so we do not include it in our model. 

\section{Discussion}\label{sec:Discussion}

\begin{figure}
  \includegraphics[width=\columnwidth]{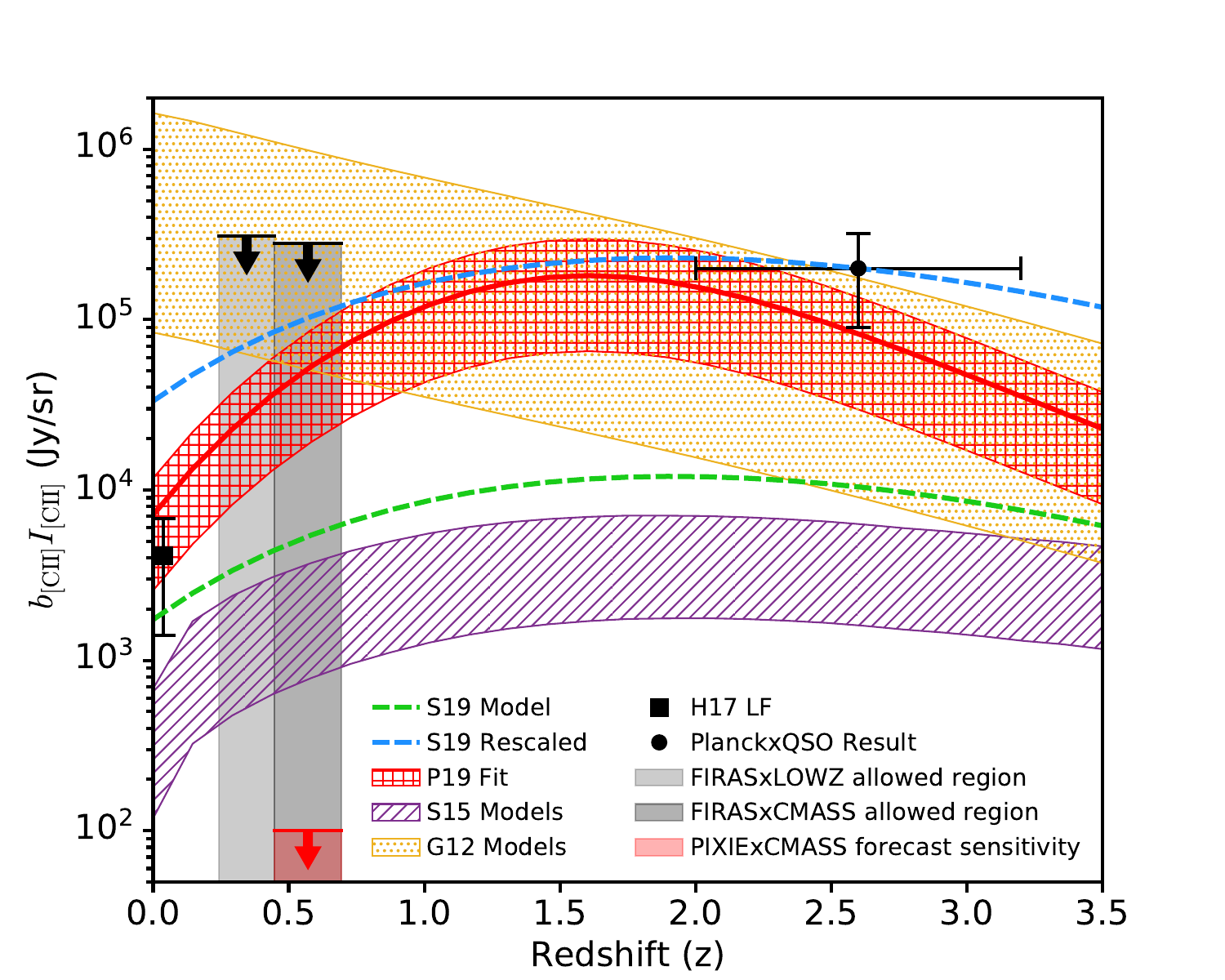}
  \caption{Comparison of measured values of $b_{\rm \cii}I_{\rm \cii}$ to various models from the literature.  The gray boxes show the values allowed by the \FLZ\, and \FC\, 95\% upper limits, with their widths representing the redshift ranges used. The black circle shows the \PB\, measurement from \citet{yang2019evidence}. The horizontal error bars show the full redshift range of that measurement, and the vertical error bars show the 95\% confidence intervals on their measurement of $b_{\rm \cii}I_{\rm \cii}$. The black square shows the value inferred from the $z\sim0$ \cii\ luminosity function \citep{2017ApJ...834...36H}, assuming $b_{\rm \cii}=1$. The error bars come from their 1-$\sigma$ errors on luminosity function parameters. 
  The purple hatched region shows the range of brightnesses allowed by the scaling-relation models from \citet{Silva2015}, the yellow dotted region shows the range of brightnesses of the collisional excitation models from \citet{Gong2012}, and the green dashed curve shows the semi-empirical model from \citet{Sun2019}.  The blue dashed curve shows the \citet{Sun2019} model linearly rescaled to match the \PB\, measurement at the appropriate redshift. The hatched red region shows the 1-$\sigma$ range for the abundance-matching fit of \citet{2019MNRAS.488.3014P} to \citet{2017ApJ...834...36H} and \PB, and the solid red line is the best-fit model. 
  The red region below 100\,Jy/sr in the \FC\, redshift range shows the projected sensitivity achievable with a cross-correlation between PIXIE and CMASS galaxies in the same redshift range used for the \FC\, analysis. This would easily detect even the most pessimistic of the plotted \cii\ models at high sensitivity.}
  \label{fig:Ib_models}
\end{figure}

Figure\,\ref{fig:Ib_models} compares \FB\ upper limits on $b_{\rm \cii}I_{\rm \cii}$ to several representative physical models of \cii\ brightness as a function of redshift. 
It also shows the \PB\ intensity mapping constraint from \cite{yang2019evidence}, assuming that the excess power detected is \cii\ emission.
Models that do not scale \cii\ emission with star formation rate evolution are disfavored because their low-redshift \cii\ emission is too bright. Models that scale \cii\ emission with star formation rate can be calibrated with physical parameters to be consistent with both \FB\ and \PB. 

The yellow dotted region in Figure\,\ref{fig:Ib_models} shows the range of \cii\ amplitudes predicted by the collisional excitation model from \citet{Gong2012}, hereafter G12.  In this model, the mean \cii\ intensity is computed through a simple radiative transfer model whose free parameters are the number density $n_e$ and kinetic temperature $T_e^K$ of electrons within the emitting galaxies.  The yellow dotted region spans the range of $T_e^K$ and $n_e$ values considered by G12.  Because this model predicts comparatively bright \cii\ emission, \citet{yang2019evidence} argue that it provides the best-fit to their measurement and use their result to place constraints on the two free parameters. \FB\ upper limits at $z {\sim} 0.5$ rule out the brightest range of the G12 predictions.  Since it is this bright end that is consistent with the \citet{yang2019evidence} measurement at $z\sim2.6$, the G12 model is disfavored to explain the \FB\ and \PB\ \cii\ measurements. 

The G12 model was originally created to forecast emission at much higher redshifts, during the epoch of reionization. Since \cii\ luminosity is expected to be correlated with star formation, we qualitatively expect the \cii\ amplitude to follow the cosmic star formation history and come to a peak around $z\sim2-3$, declining to the present day \citep{Madau2014}.  This is not seen in the G12 model, which monotonically increases as we move to lower redshift.  Thus the bright G12 predictions, which are disfavored by the combination of \FB\ and \PB, may not be a physically accurate estimate of the true \cii\ evolution during more recent epochs since $z{\sim} 2$.

Next, we consider the scaling models from \citet{Silva2015} (hereafter S15), shown as the hatched purple region in Figure\,\ref{fig:Ib_models}.  
\cite{Silva2015} consider four different empirically-calibrated power-law scaling relations between \cii\ luminosity and star formation rate (SFR).  
The hatched purple region shows the full range of \cii\ luminosity predictions from the four values of the slope and amplitude of this power-law scaling from their Table 1. Although these models were originally constructed to predict \cii\ intensity at reionization redshifts, they are correlated with the cosmic star formation history and thus show the qualitative redshift evolution we expect, with a peak at $z\sim2$ and a decline at lower redshifts. 
These predictions fall a factor of 100 or more below our \FB\ upper limits at $z\sim0.5$, putting them well below our ability to constrain.  However, these models also fall well below the \PB\, estimate, meaning they may also be pessimistic predictions.

Next, we examine the ``semi-empirical" model from \citet{Sun2019} (hereafter S19), plotted as a dashed green curve.  This calculation falls somewhat between the other two. It uses a physically motivated scaling between the \cii\ and infrared luminosity of a halo. This scaling is parameterized in terms of the photoelectric heating efficiency from dust grains, $\epsilon_{\rm{PE}}$.  The distribution of galaxy infrared luminosities is calibrated to empirical measurements of the cosmic infrared background \citep{Planck2014}. This model also predicts a fiducial \cii\ amplitude well below our limits and the \citet{yang2019evidence} measurement. However, as shown in Figure\,10 of S19, the \cii\ intensity can be brought into agreement with the \PB\, measurement, if one increases $\epsilon_{\rm{PE}}$ by a factor of ${\sim} 6$ from their fiducial number. The authors, however, note that this higher value may lead to tension with the observed relation between \cii\ luminosity and star formation rate in low-redshift galaxies.
The redshift evolution of this rescaled S19 model is plotted as the dashed blue curve in Figure\,\ref{fig:Ib_models}.  In contrast with the G12 model, we find that our FIRAS measurements are fully consistent with the redshift evolution seen here, even scaling to the brighter emission at $z{\sim}2.6$ suggested by \citet{yang2019evidence}.

We also include, in red, an empirical fit from \cite{2019MNRAS.488.3014P} (hereafter P19), which uses the technique of abundance-matching to fit a power-law luminosity function with an exponential cut-off to the measured $z {\sim} 0$ \cii\ luminosity function from \cite{2017ApJ...834...36H}. The amplitude of \cii\ emission in P19 evolves with redshift as a power-law of the observed star-formation rate density, whose redshift evolution was measured by \cite{Madau2014}. The power-law index is fit to be consistent with the $I_{\rm \cii}$ value reported by \cite{pullen2018search} at $z\sim2.6$. In order to accommodate this large increase in intensity from $z\sim 0$ to $z\sim2.6$, the best-fit value for this index is $1.79 \pm 0.30$. This is substantially higher than the values near unity assumed in S15 \citep{Silva2015}, based on observations of individual local and high-redshift galaxies. The hatched red region shows the $1-\sigma$ range for the P19 model fits, and the solid red line is the best-fit model. Because the P19 fit finds a lower mass range for \cii\ emitters than that assumed in the model of \cite{pullen2018search}, it has a lower value for $b_{\rm \cii}$, and consequently, a lower value for $b_{\rm \cii}I_{\rm \cii}$ at $z\sim2.6$.

Figure \ref{fig:Ib_models} also includes a point at $z {\sim} 0$ from the luminosity function measurements of \cite{2017ApJ...834...36H}, which integrates to $4.1\pm 2.7$\,kJy/sr. The measured galaxies cover the $10^{6.5}$ to $10^{9.5}$ solar luminosity range, via a mix of direct detection of \cii\ emission, at the bright end, and inference of \cii\ emission via FIR brightness and color measurements, at the dim end. An integral of this luminosity function yields the expected \cii\ brightness, $I_{\rm \cii}$, but in order to plot $b_{\rm \cii}I_{\rm \cii}$, a model for the \cii\ bias must be assumed. We estimate a bias of 1, which corresponds to the value from the halo model used in appendix \ref{sec:Appendix_A}. This roughly matches the bias evolution of all the models plotted, which predict decreasing bias at low redshift. This convergence of bias toward unity at low redshift occurs because, as halo masses increase at low redshift, the mass range for \cii-hosting galaxies becomes less biased towards the higher mass end of the halo distribution.

The models that calibrate the \cii-SFR relationship based on observations of known galaxies (S15) or on simple physical mechanisms (S19) struggle to decrease quickly enough at low redshift to be consistent with both \PB\ and \cite{2017ApJ...834...36H}. Assuming that the \PB\ excess is \cii\ emission, there are still several possible explanations. The estimated $z {\sim} 0$ \cii\ bias of 1 could be too low. However, the required bias of 8 to bring \cite{2017ApJ...834...36H} to agreement with the re-scaled S19 model would be extremely large compared to biases typically inferred from these \cii\ models. Alternatively, there could be an undetected population of low-luminosity \cii\ galaxies. In fact, \cite{2017ApJ...834...36H} suggest the possibility of unmeasured faint low-metallicity dwarf-galaxies, which would have a different \cii-FIR calibration, contributing to the faint end of the \cii\ luminosity function. This possibility highlights a trend where early intensity mapping results suggest unexpectedly bright cumulative emission compared to direct detection of individual objects. For instance, \cite{2021arXiv210614904B} suggests that potential bright CO emission detected by mmIME \citep{Keating_2020} may be due to a large population of dim CO galaxies. Indeed, this CO excess is of a similar order of magnitude to the difference seen here between the S19 and \PB\ \cii\ values, hinting at a possibly related cause. Another possibility is that the assumption of these models, that \cii\ emission is directly tied to the star formation rate, is too simple.

Conclusions here are limited by thermal noise in FIRAS. To meaningfully constrain the space of models, qualitatively new improvements in sensitivity are required. To put FIRAS in the context of what could be achieved by a modern mission, we compute the potential sensitivity achievable by cross-correlating maps from the proposed PIXIE satellite with the CMASS galaxy maps used in this paper. The Primordial Inflation Explorer (PIXIE) is a proposed NASA Explorer-class mission to measure the polarized imprint of primordial inflation on the CMB \citep{kogut2011primordial}, along with several other spectral distortions of the CMB \citep{chluba2021new}. PIXIE will conduct an all-sky survey with angular and spectral resolution and spectral coverage similar to FIRAS. However, the beam will be slightly smaller, and the noise level will be three orders of magnitude lower than FIRAS. These properties make PIXIE well-placed to study star-formation and scale-dependent bias as a test of non-Gaussianity via intensity mapping \citep{dizgah2019probing, switzer2017tracing}. 

To assess PIXIE's constraining power, we conduct a suite of 10,000 simulations with no \cii\ signal. These simulations project that a cross-correlation of PIXIE maps with CMASS galaxies could place an upper limit of $100$ Jy/sr on $b_{\rm \cii}I_{\rm \cii}$ at $95\%$ confidence (see the red transparent region in Figure\,\ref{fig:Ib_models}). This result means that \PIB\, is easily sensitive enough to detect even the most pessimistic \cii\ models at high significance. 

In addition to PIXIE, many of the other planned or in-progress intensity mapping surveys described in Section\,\ref{sec:introduction} could shed useful light on the redshfit evolution of the \cii\ line.  The TIM and EXCLAIM surveys will produce maps of \cii\ at redshifts $\sim1$ and $\sim3$ respectively, with considerably more sensitivity than FIRAS and better spectral resolution than Planck.  Parallel measurements of CO emission over similar redshift range, though they track somewhat different ISM physics, are still strongly correlated with star formation and will be quite complementary to the \cii\ measurments.  The COMAP survey has produced early science limits on CO(1-0) emission in a similar redshift range to the \PB\ \cii\ observation, and should have sufficient sensitivity to explore these discrepancies with their full data set \citep{Chung2021,Breysse2021}.  Similarly, though the TIME, CONCERTO, and FYST surveys primarily target reionization-era lines, they have several CO ``foregrounds" which would overlap with the FIRAS redshift range, potentially enabling deeper measurements.

\section{Conclusion}
\label{sec:Conclusion}

Our analysis constrains the product of the bias and specific intensity of \cii, $b_{\rm \cii}I_{\rm \cii}$, to be ${<}0.31$\,MJy/sr at $z{\sim}0.35$ and ${<}0.28$\,MJy/sr at $z{\sim}0.57$ at $95\%$ confidence. Through both FIRAS' unique capability for tomographic measurement and the depth of BOSS data at these redshifts, bounds on \cii\ developed here are competitive with those achieved in the recent analysis of Planck data \citep{yang2019evidence, pullen2018search}. We rule out a swath of collisional excitation models consistent with \PB, but note that those models were not designed to predict low redshift \cii. \cii\ emission is expected to correlate with the star formation rate, which appears to have peaked at $z \sim 2 - 3$ and then declined at later redshifts \citep{Madau2014}. Our constraints are consistent with all models that include this expected correlation of \cii\ emission with star formation rate. If the high \cii\ intensity measured from \PB\ is accurate, then \cii\ models with realistic redshift evolution tied to the star formation rate predict $b_{\rm \cii}I_{\rm \cii} \approx 0.1$\ MJy/sr at $z\sim0.5$. This expected value is nearly within reach of the FIRAS data. Measurements described here are limited by FIRAS thermal noise and angular resolution rather than the depth of the galaxy redshift survey, so substantial further progress will require new intensity mapping measurements. We show that the proposed PIXIE mission could detect even pessimistic models for \cii\ emission. The SHT approach developed here also applies well to wide-field surveys such as HIRAX, CHIME, TianLai, HERA, and SPHEREx, both in auto-power and crosspower. The SHT is broadly well-matched to intensity mapping analysis through its ability to capture all Gaussian information directly in observed map space and to easily account for survey effects such as the curved sky and inhomogeneous noise.

\section*{Acknowledgements}
\label{sec:Acknowledgements}
We would like to thank Dale Fixsen, Nathan Miller, and Nils Odegard for their useful input regarding the FIRAS noise and beam model. Some of the results in this paper have been derived using the healpy and HEALPix package.  PCB was supported by the James Arthur Postdoctoral Fellowship.

Funding for the Sloan Digital Sky Survey IV has been provided by the Alfred P. Sloan Foundation, the U.S. Department of Energy Office of Science, and the Participating Institutions. SDSS-IV acknowledges support and resources from the Center for High Performance Computing  at the University of Utah. The SDSS website is www.sdss.org.

SDSS-IV is managed by the Astrophysical Research Consortium for the Participating Institutions of the SDSS Collaboration including the Brazilian Participation Group, the Carnegie Institution for Science, Carnegie Mellon University, Center for Astrophysics | Harvard \& Smithsonian, the Chilean Participation Group, the French Participation Group, Instituto de Astrof\'isica de Canarias, The Johns Hopkins University, Kavli Institute for the Physics and Mathematics of the Universe (IPMU) / University of Tokyo, the Korean Participation Group, Lawrence Berkeley National Laboratory, Leibniz Institut f\"ur Astrophysik Potsdam (AIP),  Max-Planck-Institut f\"ur Astronomie (MPIA Heidelberg), Max-Planck-Institut f\"ur Astrophysik (MPA Garching), Max-Planck-Institut f\"ur Extraterrestrische Physik (MPE), National Astronomical Observatories of China, New Mexico State University, New York University, University of Notre Dame, Observat\'ario Nacional / MCTI, The Ohio State University, Pennsylvania State University, Shanghai Astronomical Observatory, United Kingdom Participation Group, Universidad Nacional Aut\'onoma de M\'exico, University of Arizona, University of Colorado Boulder, University of Oxford, University of Portsmouth, University of Utah, University of Virginia, University of Washington, University of Wisconsin, Vanderbilt University, and Yale University.

\section*{Data Availability}
The FIRAS data underlying these results are available on NASA's LAMBDA archive at \url{https://lambda.gsfc.nasa.gov/product/cobe/firas_prod_table.cfm}. The BOSS data underlying these results are on the BOSS data release 12 archive at \url{https://www.sdss.org/dr12/}.

\onecolumn
\appendix

\section{Cross-power Approximate Sensitivity Forecast}
\label{sec:appendix_sensitivity_forecast}

The complete tomographic likelihood for $C_b^{\times}(z,z')$ has a complex structure. This appendix simplifies the sensitivity to its essential elements to provide some intuition.
For a survey where the intensity map auto-power is noise and/or foreground dominated, the covariance on the cross-power along the angular diagonal ($b=b'$) is roughly
\begin{equation}\label{eq:cross_power_cov_approx_appendix}
    \langle \Delta C_b^{\times}(z_1,z_2) \Delta C_b^{\times}(z_3,z_4) \rangle \approx  \frac{1}{f_{\rm sky} \Delta \ell (2\ell + 1)} [ C_b^{\rm IM}(z_1,z_3) C_b^{g}(z_2,z_4)].
\end{equation}

FIRAS lacks the redshift resolution for there to be a significant signal in the \cii\, or galaxy angular cross-power when $z \neq z'$. Considering only cross-correlations between the same redshift slice, the covariance $\langle \Delta C_b^{\times}(z_1,z_2) \Delta C_b^{\times}(z_3,z_4) \rangle$ is zero except along the redshift diagonal $z_1 {=} z_2 {=} z_3 {=} z_4$. We also assume that the bandpowers are broad enough to have negligible correlations for $b \neq b'$. 
Dropping redshift subscripts, the binned \cii\ cross-power signal can be modeled as $C^{\times}_b = b_{\rm \cii}I_{\rm \cii}C^{\delta}_b$, where $C^{\delta}_b$ is the binned dark matter angular power spectrum, $b_{\rm \cii}$ is the bias of \cii\ galaxies, and $I_{\rm \cii}$ is the specific intensity of \cii\ emission. From this signal model and Equation\,\ref{eq:cross_power_cov_approx_appendix}, the variance on a determination of $b_{\rm \cii}I_{\rm \cii}$ from the angular cross-power of any single redshift and an angular bin is \footnote{An interesting feature of this formula is that, for a wide range of redshift bin sizes, this variance on the line amplitude is roughly the same. For instance, if FIRAS had half of its actual redshift resolution, the cosmological and galaxy clustering power spectrum would wash out by a factor of 1/2, but the increased bandwidth would also decrease the thermal noise signal in the FIRAS auto-power spectrum by 1/2. These effects would cancel out to keep the variance per angular bin and redshift slice constant.}
\begin{equation}\label{eq: bI_var_per_z_bin}
\sigma^2_{bI} |_{\rm per-z,b} \approx \frac{1}{f_{\rm sky} \Delta \ell (2\ell + 1)} \frac{C_b^{\rm IM} C_b^{g}}{C^{\delta}_b C^{\delta}_b}. 
\end{equation}
The total constraining power is an inverse variance weighted sum over the angular and redshift bins,
\begin{equation}\label{eq:bI_var}
\sigma^2_{bI} \approx \left[ \sum_{b,z} \left( \frac{1}{f_{\rm sky} \Delta \ell (2\ell + 1)} \frac{ C_b^{\rm IM} C_b^{g}}{C^{\delta}_bC^{\delta}_b} \right)^{-1} \right]^{-1}. 
\end{equation}
From this formula, we predict a sensitivity of 0.27\,MJy/sr ($2\sigma$) on $b_{\rm \cii}I_{\rm \cii}$. In computing this estimate, we have neglected foregrounds and assumed that only thermal noise contributes to $C_b^{\rm IM}$, since smooth foregrounds are de-weighted with only a small penalty to sensitivity. This simple estimate closely matches the precision we achieve with the complete calculation. 

\section{Details of model for \texorpdfstring{\FB}{FIRASxBOSS}}
\label{sec:Appendix_A}

To model the dust and \cii\ emission, we employ a technique similar to \cite{pullen2018search}.  Using the halo model, the redshift distribution of 
specific intensity is 
\begin{equation}\label{eq:dI/dz}
\begin{split}
\frac{dI_{\nu}}{dz} = \frac{d\chi}{dz} \frac{dI_{\nu}}{d\chi} &= \frac{d\chi}{dz} \int dM 
\frac{L_{\nu}(M,z) }{4\pi}\frac{dn(z)}{dM}\\
&= \frac{c}{(1+z)H(z)}\mean{j_{\nu}}(z),
\end{split}
\end{equation}
where $L_{\nu}(M,z)$ represents the redshifted specific luminosity for a halo of mass $M$, given by $L_{\nu}(M,z) = \frac{L_{\nu(1+z)}(M,z)}{1+z}$, where $L_{\nu(1+z)}(M,z)$ represents the rest-frame specific luminosity evaluated at $\nu(1+z)$. On the second line of Equation\,\ref{eq:dI/dz}, we introduced the mean comoving emission coefficient $\mean{j_{\nu}}(z)$, defined as
\begin{equation}
\mean{j_{\nu}}(z) = \int dM \frac{L_{\nu(1+z)}(M,z)}{4\pi}\frac{dn(z)}{dM}.
\end{equation}
Following the analysis of extragalactic cosmic infrared background (CIB) emission of \cite{10.1111/j.1365-2966.2012.20510.x}, the specific luminosity is parameterized as
\begin{equation}\label{eq:shang_parametrization}
L_{\nu(1+z)}(M,z) = L_0 \Phi(z) \Sigma(M)\Theta \left[ \nu(1+z)\right],
\end{equation}
where $L_0$ is a normalization constant, $\Sigma(M)$ parameterizes the mass dependence of the halo luminosity, and $\Phi(z)$ parameterizes the redshift evolution of both the continuum and \cii\ luminosity, which we assume evolve together. $\Phi(z)$ is expected to have peaked around $z\sim2-3$ \citep{Madau2014}. A sufficiently precise intensity mapping experiment with a large frequency bandwidth can, in principle, constrain the form of the redshift dependence of $\Phi(z)$. \FB, however, lacks the precision to measure this evolution, so we choose to model it with a simple power-law, $\Phi(z) = (1+z)^{2.3}$, following the parameterization and fit of \cite{pullen2018search}. $\Sigma(M)$ is assumed to follow a log-normal distribution
\begin{equation}
\Sigma(M) \propto \frac{M}{\sqrt{2\pi\sigma^2_{L/M}}}\exp{-\left[\frac{(\log_{10}M - \log_{10}M_{\rm eff})^2}{2\sigma^2_{L/M}}\right]},
\end{equation}
where $\log_{10}(M_{\rm eff}/M_{\odot}) = 12.6$ \citep{Planck2014, serra2016dissecting} and $\sigma^2_{L/M}=0.5$ \citep{10.1111/j.1365-2966.2012.20510.x, Planck2014, serra2016dissecting}. The integral over the halo mass function $dn(z)/dM$ is computed with the python package hmf \citep{2014ascl.soft12006M}, assuming Planck 2015 \citep{Ade:2015xua} cosmological parameters and a \cite{2008ApJ...688..709T} functional form for $dn(z)/dM$.
We include \cii\ emission in this specific luminosity by assuming that each CIB galaxy emits continuum emission and a narrow \cii\ line. Assuming that the ratio of total power emitted by the line and the continuum is constant, and ignoring the finite mass-dependent velocity width of the galaxy, we parameterize $\Theta(\nu)$ as the sum of continuum emission ($\Theta^c(\nu)$, parametric form from \cite{10.1111/j.1365-2966.2012.20510.x}) and \cii\ line emission
\begin{equation}
\Theta(\nu) = \Theta^c(\nu) + \Theta^{\rm \cii}(\nu),
\end{equation}
\begin{equation*}
\Theta^c(\nu) = \left\{
  \begin{array}{@{}ll@{}}
    (\nu/\nu_0)^{\beta}[B_{\nu}(T_d)/B_{\nu_0}(T_d)] & \text{~~~~if}\ \nu<\nu_0 \\
    (\nu/\nu_0)^{-2} & \text{~~~~otherwise},
  \end{array}\right. \nonumber 
\end{equation*}
\begin{equation*}
\Theta^{\rm \cii}(\nu) = \left[\int\Theta^c(\nu')d\nu'\right]\alpha\delta(\nu -\nu_{\rm \cii}), \nonumber
\end{equation*}
where we have defined $\alpha$ to be the ratio of the total power of the continuum emission to the total power of \cii\ line emission. We choose $\beta=1.5$, in agreement with \cite{2014A&A...566A..55P}. \cite{2014A&A...570A..98S} measured the dust temperature of the BOSS CIB to be 26\,K, which means all of the CIB emission in the FIRAS frequencies of this analysis is below the turnover frequency $\nu_0$, so only the semi-thermal form of $\Theta^c(\nu)$ is used. 

The FIRAS instrument measures the average intensity in a given frequency bin. Ignoring redshift space distortions and any contributions from thermal noise and Milky Way foregrounds, the observed intensity is given by
\begin{equation}\label{eq:split intensity}
I_{\nu}(\theta, \phi) = \frac{1}{\Delta\nu}\int_{\nu - \Delta\nu/2}^{\nu + \Delta\nu/2}d\nu' \int \left[ \frac{dI^c_{\nu'}}{dz} + \frac{dI^{\rm \cii}_{\nu'}}{dz} \right] \cdot 
\left[1 + b_{\rm CIB}\delta\left( \chi \left( z \right),\theta,\phi \right) \right]dz,
\end{equation}
where we have explicitly separated the continuum and \cii\ contributions. We do this because the cross-correlation signal between FIRAS and BOSS has two components: (1) a galaxy-\cii\ cross-correlation component with narrow redshift window functions for both the optical galaxies and the \cii\ sources, and (2) a galaxy-continuum component with a narrow redshift window function for the optical galaxies but a broad redshift window function, given by $\frac{dI^c_{\nu}}{dz}$, for the CIB continuum:
\begin{equation}
C_{\ell}^{\times}(z,z') = C_{\ell}^{\rm \cii \times g}(z,z') + C_{\ell}^{c \times g}(z,z').
\end{equation}
The angular power spectrum between signal types $i$ and $j$ is given by
\begin{equation}
C_{\ell}^{i \times j}(z,z') = \frac{2}{\pi}\int W^i_z(k)W^j_{z'}(k)k^2 P(k,z=0) dk,
\end{equation}
where (ignoring RSD for now, and using equations \ref{eq:dI/dz} through \ref{eq:split intensity} for the \cii\ and CIB signals) the Window functions for optical galaxies, \cii, and continuum emission are as follows:
\begin{equation}
W^g_z(k) = \frac{1}{\Delta z} \int_{z_{\rm min}}^{z_{\rm max}}b_gG(z',k)j_{\ell}(k\chi(z'))dz', 
\end{equation}
\begin{equation*}
W^{\rm \cii}_z(k) = \left[\int\Theta^c(\nu)d\nu\right]\frac{\alpha}{\Delta \nu}\int_{z_{\rm min}}^{z_{\rm max}}b_{\rm \cii}G(z',k)j_{\ell}
(k\chi(z')) \frac{cL_0\Phi(z')}{4\pi(1+z')^2H(z')}\int\Sigma(M)\frac{dn(z')}{dM}dMdz', 
\end{equation*}
\begin{equation*}
W^c_z(k) = \int_{\nu - \Delta\nu/2}^{\nu + \Delta\nu/2} \frac{d{\nu'}}{\Delta \nu} \int_0^{\infty} \frac{cL_0\Phi(z')b_{\rm \cii}G(z',k)j_{\ell}
(k\chi(z'))}{4\pi(1+z')H(z')} \Theta^c(\nu'(1+z'))\int \Sigma(M)\frac{dn(z')}{dM}dMdz', 
\end{equation*}
where $G(z,k)$ is the growth factor and $z_{\rm min}$ and $z_{\rm max}$ are set by the spectral channels of FIRAS for $W^{\rm \cii}_z(k)$. As stated in Section\,\ref{sec:BOSS}, the galaxy window function $W^g_z(k)$ was chosen to have the same $z_{\rm min}$ and $z_{\rm max}$ as the FIRAS spectral channels.

If we make the reasonable approximation that all $z$-dependent quantities of the dust and \cii\ model vary slowly within the redshift bins determined by the FIRAS frequency channels, then we can compute $C_{\ell}^{\rm \cii \times g}(z,z')$ and $C_{\ell}^{c \times g}(z,z')$ fully, including linear redshift space distortions, in terms of $C^{(2)}_{\ell}(z,z')$, $C^{(1)}_{\ell}(z,z')$, and $C^{(0)}_{\ell}(z,z')$, which were defined in Section\,\ref{subsec:dark_matter_model}. For $C_{\ell}^{\rm \cii \times g}(z,z')$, we have
\begin{equation}\label{eq:cii_cl_full}
C_{\ell}^{\rm \cii \times g}(z,z') = I_{\rm \cii}(z) \cdot \left [ b_gb_{\rm \cii}C^{(2)}_{\ell}(z,z') + \frac{b_g + b_{\rm \cii}}{2}C^{(1)}_{\ell}(z,z') + C^{(0)}_{\ell}(z,z')  \right ], 
\end{equation}
where
\begin{equation}\label{eq:I_CII}
I_{\rm \cii}(z) = \left[\int\Theta^c(\nu')d\nu'\right]\frac{\alpha}{ \nu_{\rm \cii}}\frac{cL_0\Phi(z)}{4\pi H(z)}\int\Sigma(M)\frac{dn(z)}{dM}dM.
\end{equation}
In Equation\,\ref{eq:I_CII}, we have used $\Delta \nu = \frac{\nu_{\rm \cii} \Delta z}{(1+z)^2}$.
The \cii\ intensity $I_{\rm \cii}(z)$ has a redshift dependence, determined by both cosmological factors $\left( \frac{dn(z)/dM}{H(z)}\right)$ and the assumed specific luminosity parameterization $\left( L_0 \Phi(z) \Sigma(M) \right)$ of Equation\,\ref{eq:shang_parametrization}. They lead to a \cii\ intensity that increases by a factor of about 20 percent over each of the CMASS and LOWZ redshift ranges. In our analysis, the shape of this redshift evolution is fixed, and we fit only for the overall amplitude of $b_{\rm \cii}I_{\rm \cii}(z=z_{\rm center})$. 
We calculate the CIB signal in a similar manner, as
\begin{equation}\label{eq:cib_cl_full}
C_{\ell}^{c \times g}(z,z') =  \sum_{z''} \frac{dI_{\rm CIB}(\nu_{\rm \cii}^z,z'')}{dz''}\Delta z'' \cdot \left [ b_g b_{\rm \cii}C^{(2)}_{\ell}(z'',z') + \frac{b_g + b_{\rm \cii}}{2}C^{(1)}_{\ell}(z'',z') + C^{(0)}_{\ell}(z'',z')  \right ],
\end{equation}
where $\frac{dI_{\rm CIB}(\nu_{\rm \cii}^z,z'')}{dz''}\Delta z''$ is the intensity of the CIB that is emitted from sources in a redshift bin of size $\Delta z''$, centered at redshift $z''$, and measured at a frequency of $\nu_{\rm \cii}/(1+z)$. Assuming the CIB spectrum does not change significantly within the FIRAS frequency bin, then $\frac{dI_{\rm CIB}(\nu_{\rm \cii}^z,z'')}{dz''}\Delta z''$ is
\begin{equation}
\frac{dI_{\rm CIB}(\nu_{\rm \cii}^z,z'')}{dz''}\Delta z'' =  \Delta z'' \cdot  \left[ \Theta^c \left(\nu_{\rm \cii}\frac{1+z''}{1+z}\right) \frac{cL_0\Phi(z'')}{4\pi(1+z'')H(z'')}\int\Sigma(M)\frac{dn(z'')}{dM}dM \right].
\end{equation}
The sum over $z''$ in Equation\,\ref{eq:cib_cl_full} should, in principle, be carried out over all redshifts, even those outside of the galaxy survey. In practice, for the redshift bins and $\ell$ bins considered in this analysis, the bracketed $C_{\ell}(z'',z')$ kernel is dominated by the $z''= z'$ term, with the neighboring off-diagonal terms around 20\% of the diagonal term, and the rest of the terms are negligible. Analogously to our \cii\ analysis, we do not attempt to constrain the redshift evolution of the CIB brightness or the spectral shape of the CIB emission. Instead, these are fixed by the assumed values of $\beta = 1.5$, $T_d=26$\,K, $\Phi(z) = (1+z)^{2.3}$, $\log_{10}(M_{\rm eff}/M_{\odot}) = 12.6$ , and $\sigma^2_{L/M}=0.5$. Because of this assumed spectral and redshift evolution, the $\frac{dI_{\rm CIB}}{dz}$ parameter that we fit for is only the exact continuum intensity per unit redshift at $z_{\rm center}$, the central redshift of the survey, and $\nu_{\rm center}$, the central frequency of the survey. 

Equations \ref{eq:cii_cl_full} and \ref{eq:cib_cl_full} show how high spectral resolution allows a separation of correlated line and continuum emission in intensity mapping. 
In the limit where the spectral resolution is very low and $\Delta z''$ is high, all the off-diagonal terms of $C_{\ell}(z'',z')$ are zero, collapsing the sum of Equation\,\ref{eq:cib_cl_full} to only the $z''=z'$ term. Dividing the \cii\ signal by the CIB signal, one finds that the ratio of \cii\ to CIB cross-power scales as $\frac{1}{\Delta z'}$ (this point is subtle because one is free to choose $\Delta z''$ to be an arbitrarily small value in Equation\,\ref{eq:cib_cl_full}, but the redshift window for Equation\,\ref{eq:cii_cl_full} is fixed by the spectral resolution of the intensity mapping survey. Therefore, one can choose the same redshift window for the CIB calculation so that the $C_{\ell}(z,z')$ terms cancel.). This simple $\frac{1}{\Delta z'}$ scaling, where higher redshift resolution continues to linearly improve the ability to distinguish line emission from continuum emission, remains true until the redshift resolution is high enough to resolve off-diagonal components of $C_{\ell}(z'',z')$.

\section{Covariance angular coupling approximation}
\label{sec:Appendix_coupling_approximation}
This section reviews a formula for approximating the $b-b'$ coupling in the $C_b$ covariance. The formula comes directly from the work of \cite{tristram2005xspect}, originally developed for \pcl\ cross-correlation studies of CMB maps from different instruments or different detectors of the same instrument. We use the notation of Section\,\ref{sec:C_ell_Analysis}, where unbinned $C_{\ell}$ quantities retain the $\ell$ subscript, and binned quantities are written with the subscript $b$. Let $C_b^{AB}$ represent the binned angular power spectrum between maps of type $A$ and type $B$, where $A$ and $B$ can represent different redshifts or different types of maps (intensity maps or binned galaxy surveys). Then, the covariance can be written

\begin{equation}\label{eq:Tristram_cl_cov}
\langle \Delta C_b^{AB} C_{b'}^{CD}\rangle    = \sum_{b_1, b_2} (M^{AB}_{b b_{1}})^{-1} \langle \Delta D_{b_1}^{AB} \Delta D_{b_2}^{CD}\rangle (M^{CD}_{b' b_{2}})^{-1},
\end{equation}
where $(M^{AB}_{b b_{1}})^{-1}$ represents the inverse of the binned mixing matrix between maps $A$ and $B$, defined by equations \ref{eq:mixing_matrix} and \ref{eq:binning_equation}, and where $\langle \Delta D_{b_1}^{AB} \Delta D_{b_2}^{CD}\rangle$ is the covariance of the \pcl, given by
\begin{equation}\label{eq:Tristram_pcl_cov}
\langle \Delta D_{b_1}^{AB} \Delta D_{b_2}^{CD}\rangle = \sum_{\ell_1, \ell_2} B_{b_1, \ell_1}B_{b_2, \ell_2}\frac{\omega^A_{\ell_1}\omega^B_{\ell_1}\omega^C_{\ell_2}\omega^D_{\ell_2}}{2\ell_2 + 1} [M^{(2)}_{\ell_1,\ell_2}(W^{AC,BD})C_{\ell_1}^{AC}C_{\ell_2}^{BD} + M^{(2)}_{\ell_1,\ell_2}(W^{AD,BC})C_{\ell_1}^{AD}C_{\ell_2}^{BC}].
\end{equation}
In the above formula, $B_{b_1, \ell_1}$ is a binning matrix, $\omega^A_{\ell_1}$ is the combined beam and pixel window function for map $A$ (and analogously for maps $B$, $C$ and $D$), and $M^{(2)}_{\ell_1,\ell_2}(W^{AC,BD})$ is the cross-correlation matrix, defined by 
\begin{equation}
M^{(2)}_{\ell_1,\ell_2}(W^{AC,BD}) = \frac{2\ell_2 + 1}{4 \pi} \sum_{\ell_3}(2\ell_3+1)W^{AC,BD}_{\ell_3}\begin{pmatrix} \ell_1&\ell_2&\ell_3 \\ 0&0&0\end{pmatrix}^2,
\end{equation}
where
\begin{equation}
W^{AC,BD}_{\ell} = \frac{1}{2\ell + 1} \sum_m w^A_{\ell,m}w^C_{\ell,m} (w^B_{\ell,m}w^D_{\ell,m})^*.
\end{equation}
In the above formula, $w^A_{\ell,m}$ represents a spherical harmonic coefficient of the inverse noise weights for map $A$. 

The covariance presented in this section uses an approximation that works best when the sky coverage is large. We find it sufficient for our \BB\ analysis, but for \FF\ and \FB, the formula appears to be inaccurate, possibly due to the combination of the steep angular index on the foregrounds combined with the small sky fraction. We instead use a simulated covariance for \FF\ and \FB.

\section{Updating the covariance for \texorpdfstring{\FB}{FIRASxBOSS}}
\label{sec:Appendix_B}
We calculate a simulated covariance by drawing individual $a_{\ell m}$ amplitudes from our FIRAS foregrounds plus noise model, cosmological signal model, and galaxy shot noise model. Each of these models is assumed to be Gaussian and uncorrelated. We then compute simulated full-sky galaxy maps from the sum of the cosmological and shot noise draws. We also compute simulated full-sky FIRAS maps from the sum of the foreground and noise realization plus a painting of \cii\ and CIB signal onto the cosmological signal realization. The simulated FIRAS signal is then convolved with the FIRAS beam, scan, and pixel window function. Then, the simulated FIRAS and BOSS maps are multiplied by their respective weights, and the cross-power spectrum between the FIRAS and BOSS maps is computed. A simulated covariance is calculated from 10,000 of these realizations. This procedure assumes a specific value for the \cii\ and CIB brightness, but the covariance should change as the \cii\ and CIB magnitude change. So, we use four sets of simulations to create a simulated covariance that correctly updates its \cii\ and CIB contribution to match the current \cii\ and CIB magnitude estimates.

From the assumption of Gaussian power spectra, the rough form of the \FB\, covariance is
\begin{equation}
\label{eq:gaussian_cross_variance}
    \langle \Delta C_b^{\times}(z_1,z_2) \Delta C_b^{\times}(z_3,z_4) \rangle \approx  \frac{1}{f_{\rm sky} \Delta \ell (2\ell + 1)} [ C_b^{\rm IM}(z_1,z_3) C_b^{g}(z_2,z_4) +
    C_b^{\times}(z_1,z_4) C_b^{\times}(z_3,z_2)],
\end{equation}
where the partial sky also introduces some level of coupling between different angular bins, which we compute via our simulations. Schematically, dropping the redshift indices, the two terms in brackets can be broken down as follows.
\begin{equation}
\label{eq:gaussian_cross_variance_expanded_auto_terms}
    C_b^{\rm IM}C_b^{g} \propto [ C_b^{FG+N} + C_b^{\rm \cii\times \cii} + C_b^{\rm CIB\times CIB} + C_b^{\rm \cii\times CIB} + C_b^{\rm CIB \times \cii}]C_b^{g}.
\end{equation}
\begin{equation}
\label{eq:gaussian_cross_variance_expanded_cross_terms}
    C_b^{\times}C_b^{\times} \propto [C_b^{\rm \cii\times g} + C_b^{\rm CIB \times g}][C_b^{\rm \cii\times g} + C_b^{\rm CIB \times g}],
\end{equation}
where $FG+N$ refers to the model of Milky Way foregrounds plus noise.
Due to the large foregrounds and noise, the magnitude of the covariance is dominated by the $C_b^{FG+N}C_b^{g}$ term. However, we find that the best-fit values for $b_{\rm \cii}I_{\rm \cii}$ and $b_{\rm \cii}dI_{\rm CIB}/dz$ are somewhat affected by the values one assumes for them in the covariance model, so we need to be able to vary their magnitude in our covariance model.  This requires four simulations, which we describe in table \ref{tab:sim_table}.

\begin{table}
	\centering
	\caption{The four simulations required for computing the \FB\ covariance. All use the same \BB\ parameters, but the foreground, thermal noise, \cii, and CIB amplitudes vary.}
	\label{tab:sim_table}
\begin{tabular}{ |p{2cm}||p{3.5cm}|p{2cm}|p{2cm}|p{3.5cm}|  }
 \hline
 \multicolumn{5}{|c|}{Simulated Covariances} \\
 \hline
 Name & FG, Noise Amplitude & $b_{\rm \cii}I_{\rm \cii}$ &$b_{\rm \cii}dI_{\rm CIB}/dz$ & $b_g, A_{SN}$ \\
 \hline
 $\Sigma_{FG+N}$ & \FF\, best-fit & 0 & 0 & \BB\, best-fit\\
 $\Sigma_{\rm \cii}$ & 0 & 1 & 0 & \BB\, best-fit\\
 $\Sigma_{\rm CIB}$ & 0 & 0 & 1 & \BB\, best-fit\\
 $\Sigma_{\rm \cii + CIB}$ & 0 & 1 & 1 & \BB\, best-fit\\
 \hline
\end{tabular}
\end{table}

These four simulations give the following approximate covariances:
\begin{equation}
\Sigma_{FG+N} \approx \frac{1}{f_{\rm sky} \Delta \ell (2\ell + 1)} C_b^{FG+N}C_b^{g}
\end{equation}
\begin{equation}
\Sigma_{\rm \cii} \approx \frac{1}{f_{\rm sky} \Delta \ell (2\ell + 1)} [C_b^{\rm \cii\times \cii}C_b^{g} + C_b^{\rm \cii\times g}C_b^{\rm \cii\times g}]
\end{equation}
\begin{equation}
\Sigma_{\rm CIB} \approx \frac{1}{f_{\rm sky} \Delta \ell (2\ell + 1)} [C_b^{\rm CIB\times CIB}C_b^{g} + C_b^{\rm CIB\times g}C_b^{\rm CIB\times g}]
\end{equation}
\begin{equation}
\Sigma_{\rm \cii + CIB} \approx \Sigma_{\rm \cii} + \Sigma_{\rm CIB} + \frac{1}{f_{\rm sky} \Delta \ell (2\ell + 1)}\left[ \left(C_b^{\rm \cii\times CIB} + C_b^{\rm CIB \times \cii}\right)C_b^{g} + C_b^{\rm \cii\times g}C_b^{\rm CIB\times g} + C_b^{\rm CIB\times g}C_b^{\rm \cii\times g} \right]
\end{equation}
From the bottom three simulations, we can isolate the term that arises due to cross-correlations between \cii\ and CIB, as follows:
\begin{equation}
\Sigma_{\rm \cii \times CIB} = \Sigma_{\rm \cii + CIB} - \Sigma_{\rm \cii} - \Sigma_{\rm CIB}.
\end{equation}
The total covariance for arbitrary \cii\ and CIB amplitudes can now be constructed as follows:
\begin{equation}
\label{eq:update_cov}
\Sigma = \Sigma_{FG+N} +b_{\rm \cii}^2I_{\rm \cii}^2\Sigma_{\rm \cii} + b_{\rm \cii}^2(dI_{\rm CIB}/dz)^2\Sigma_{\rm CIB} + b_{\rm \cii}^2I_{\rm \cii}(dI_{\rm CIB}/dz)\Sigma_{\rm \cii \times CIB}.
\end{equation}
This is the covariance used for our cross-power estimator in Section\,\ref{subsec:cross-power} and in our FIRAS validation simulations, described in Appendix\,\ref{sec:Appendix_cov_sims}. 

\section{Validating the parametric FIRAS model in the covariance for \texorpdfstring{\FB}{FIRASxBOSS}}
\label{sec:Appendix_cov_sims}

\begin{figure}
  \includegraphics[width=6in]{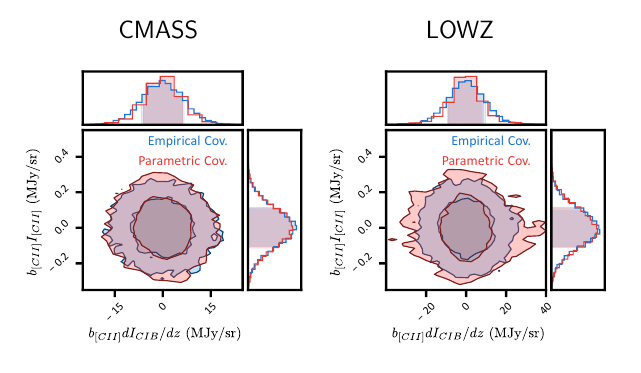}
  \caption{Simulations of fits to the CIB and \cii\ amplitude for two different types of FIRAS models.  The red contours show simulations that use the parametric FIRAS model described in Section\,\ref{subsec:FIRAS_auto}. The blue contours show simulations that use the empirical FIRAS model, where the unbinned FIRAS data are used directly as the FIRAS model. The left and right plots show the  \FC\, and \FLZ\, simulations, respectively.}
  \label{fig:cross_power_sims}
\end{figure}

We compare two sets of simulations, each performed over both the \cmass\, and \lowz\, regions. In the first set of simulations, we draw the FIRAS $a_{\ell m}$ amplitudes from the parametric model fit described in Section\,\ref{subsec:FIRAS_auto}. In the second set of simulations, the FIRAS $a_{\ell m}$ amplitudes are drawn from an empirical FIRAS model that is the measured FIRAS auto-power in bandpowers. In all cases, the BOSS signal is drawn from the parametric fit of Section\,\ref{subsec:BOSS_auto_model}, and both the \cii\ and CIB amplitudes are set to zero. We draw 10,000 realizations for each set of simulations and compute the covariance directly from the simulations. We then find the maximum likelihood solution for each realization, using a Gaussian likelihood with this simulated covariance (and also updating the covariance to account for the current \cii\ and CIB estimate according to Equation\,\ref{eq:update_cov}). Note that the empirical FIRAS-model simulations are not subject to cosmic bias because the real \cii\ and CIB signal present in the empirical FIRAS model is not correlated with the simulated cosmological clustering realization. Figure\,\ref{fig:cross_power_sims} shows that the empirical (blue) and parametric (red) FIRAS simulations lead to near-identical uncertainties on $b_{\rm \cii}I_{\rm \cii}$ and $b_{\rm \cii}dI_{\rm CIB}/dz$. 

As a consistency check on our MCMC procedure, we also perform an MCMC analysis on one of our realizations for the parametric covariance. We verify that the 95 and 68 percent MCMC contours are nearly identical in size and shape to the 95 and 68 percent contours of the histogram of maximum likelihood points from our 10,000 simulations. This calculation confirms that the parametric variance model for \FF\ yield errors consistent with those inferred from the measured \FF\ power spectrum (that the parametric \FF\ model is sufficiently accurate).
\\

\twocolumn
\bibliographystyle{mnras}
\bibliography{main}

\bsp	
\label{lastpage}
\end{document}